\documentclass[a4paper]{article}
\pdfoutput=1

\usepackage[hyphens]{url}
\usepackage[top=2cm,bottom=3cm,left=2.4cm,right=2.4cm]{geometry}
\usepackage{mathtools,amssymb}
\usepackage[usenames,dvipsnames,table]{xcolor}
\usepackage[titletoc]{appendix}
\usepackage{mathtools,amssymb}
\usepackage[makeroom]{cancel}
\usepackage{graphicx}
\usepackage{epstopdf}
\usepackage{mathrsfs}
\usepackage{enumitem}
\usepackage{stackrel}
\usepackage{float}
\usepackage{booktabs}
\usepackage[normalem]{ulem}  
\usepackage{cancel} 
\usepackage{hyperref}
\usepackage[stable]{footmisc}
\usepackage[draft,color,final]{showkeys} 
\usepackage{physics}
\usepackage{empheq}
\usepackage{cite} 
\usepackage{comment}
\usepackage{braket}
\usepackage{caption}
\usepackage{subcaption}
\usepackage{tikz}
\usetikzlibrary{decorations.pathreplacing}
\usepackage{verbatim}
\usepackage{simpler-wick}
\usepackage[T1]{fontenc}
\usepackage{esint}

\DeclareMathAlphabet{\mathpzc}{OT1}{pzc}{m}{it}

\allowdisplaybreaks
\allowbreak

\definecolor{refkey}{gray}{0.75}
\definecolor{labelkey}{RGB}{155,48,48}
\renewcommand*\showkeyslabelformat[1]{%
  \fbox{\parbox[t]{0.8\marginparwidth}{\raggedright\normalfont\scriptsize\url{#1}}}}

\usepackage{etoolbox}
\makeatletter
\patchcmd{\hyper@makecurrent}{%
    \ifx\Hy@param\Hy@chapterstring
    \let\Hy@param\Hy@chapapp
    \fi
}{%
    \iftoggle{inappendix}{
	\@checkappendixparam{chapter}%
	\@checkappendixparam{section}%
	\@checkappendixparam{subsection}%
	\@checkappendixparam{subsubsection}%
	\@checkappendixparam{paragraph}%
	\@checkappendixparam{subparagraph}%
    }{}%
}{}{ \errmessage{failed to patch}}

\newcommand*{\@checkappendixparam}[1]{%
	\def\@checkappendixparamtmp{#1}%
	\ifx\Hy@param\@checkappendixparamtmp
	\let\Hy@param\Hy@appendixstring
	\fi
}
\makeatletter

\newtoggle{inappendix}
\togglefalse{inappendix}

\apptocmd{\appendix}{\toggletrue{inappendix}}{}{\errmessage{failed to patch}}
\apptocmd{\subappendices}{\toggletrue{inappendix}}{}{\errmessage{failed to patch}}

\newcommand{\lsim}{\mathrel{\hbox{\rlap{\lower .55ex
\hbox{$\sim$}} \kern-.3em \raise.4ex \hbox{$<$}}}}
\newcommand{\gsim}{\mathrel{\hbox{\rlap{\lower.55ex
\hbox{$\sim$}} \kern-.3em \raise.4ex \hbox{$<$}}}}

\begin{document}


\newcommand{\partiald}[2]{\dfrac{\partial #1}{\partial #2}}
\newcommand{\beq}{\begin{equation}}
\newcommand{\eeq}{\end{equation}}
\newcommand{\f}{\frac}
\newcommand{\s}{\sqrt}
\newcommand{\lm}{\mathcal{L}}
\newcommand{\wm}{\mathcal{W}}
\newcommand{\om}{\mathcal{O}_{n}}
\newcommand{\bea}{\begin{eqnarray}}
\newcommand{\eea}{\end{eqnarray}}
\newcommand{\ba}{\begin{align}}
\newcommand{\ea}{\end{align}}
\newcommand{\ep}{\epsilon}

\def\gap#1{\vspace{#1 ex}}
\def\be{\begin{equation}}
\def\ee{\end{equation}}
\def\bal{\begin{array}{l}}
\def\ba#1{\begin{array}{#1}}  
\def\ea{\end{array}}
\def\bea{\begin{eqnarray}}
\def\eea{\end{eqnarray}}
\def\beas{\begin{eqnarray*}}
\def\eeas{\end{eqnarray*}}
\def\del{\partial}
\def\eq#1{(\ref{#1})}
\def\fig#1{Fig \ref{#1}} 
\def\re#1{{\bf #1}}
\def\bull{$\bullet$}
\def\nn{\nonumber}
\def\ub{\underbar}
\def\nl{\hfill\break}
\def\ni{\noindent}
\def\bibi{\bibitem}
\def\vev#1{\langle #1 \rangle} 
\def\mattwo#1#2#3#4{\left(\begin{array}{cc}#1&#2\\#3&#4\end{array}\right)} 
\def\tgen#1{T^{#1}}
\def\half{\frac12}
\def\floor#1{{\lfloor #1 \rfloor}}
\def\ceil#1{{\lceil #1 \rceil}}

\def\Tr{{\rm Tr}}

\def\mysec#1{\gap1\ni{\bf #1}\gap1}
\def\mycap#1{\begin{quote}{\footnotesize #1}\end{quote}}

\def\Red#1{{\color{red}#1}}
\def\blue#1{{\color{blue}#1}}
\def\Om{\Omega}
\def\a{\alpha}
\def\b{\beta}
\def\l{\lambda}
\def\g{\gamma}
\def\e{\epsilon}
\def\Si{\Sigma}
\def\p{\phi}
\def\z{\zeta}

\def\lan{\langle}
\def\ran{\rangle}

\def\bit{\begin{item}}
\def\eit{\end{item}}
\def\benu{\begin{enumerate}}
\def\eenu{\end{enumerate}}

\def\tr{{\rm tr}}
\def\intk#1{{\int\kern-#1pt}}


\parindent=0pt
\parskip = 10pt

\def\al{\alpha}
\def\ga{\gamma}
\def\Ga{\Gamma}
\def\G{\Gamma}
\def\be{\beta}
\def\de{\delta}
\def\De{\Delta}
\def\ep{\epsilon}
\def\ro{\rho}
\def\la{\lambda}
\def\La{\Lambda}
\def\ka{\kappa}
\def\om{\omega}
\def\si{\sigma}
\def\th{\theta}
\def\ze{\zeta}
\def\ne{\eta}
\def\del{\partial}
\def\cdev{\nabla}

\def\gh{\hat{g}}
\def\Rh{\hat{R}}
\def\Boxh{\hat{\Box}}
\def\Kb{\mathcal{K}}
\def\phit{\tilde{\phi}}
\def\gt{\tilde{g}}
\newcommand{\h}{\hat}
\newcommand{\ti}{\tilde}
\newcommand{\sD}{{\mathcal{D}}}
\newcommand{\colored}[1]{ {\color{turquoise} #1 } }
\newcommand{\propBbd}{\mathcal{G}}
\newcommand{\propBB}{\mathbb{G}}
\newcommand{\christof}[3]{ {\Ga^{#1}}_{#2 #3}}
\def\ads{AdS$_{\text{2}}$~}
\def\GN{G$_{\text{N}}$~}
\def\zb{{\bar{z}}}
\def\fb{\bar{f}}
\def\delb{\bar{\del}}
\def\wb{\bar{w}}
\def\gb{\bar{g}}
\def\gp{g_+}
\def\gm{g_-}
\def\phit{\tilde{\phi}}
\def\xb{\bar{x}}
\def\yb{\bar{y}}
\def\xp{x_+}
\def\xm{x_-}
\def\finv{\mathfrak{f}_i}
\def\fbinv{\bar{\mathfrak{f}}_i}
\def\gc{\mathfrak{g}}
\def\gcb{\bar{ \mathfrak{g}}}
\def\disc{\mathcal{D}}
\def\rhp{\mathbb{H}}
\def\picklemma{Schwarz-Pick lemma}
\def\mobius{M\"{o}bius~}
\def\ft{\tilde{f}}
\def\zet{\tilde{\ze}}
\def\taut{\tilde{\tau}}
\def\thet{\tilde{\theta}}
\def\slr{\ensuremath{\mathbb{SL}(2,\mathbb{R})}}
\def\slc{\ensuremath{\mathbb{SL}(2,\mathbb{C})}}
\def\nh{\hat{n}}
\def\cD{\mathcal{D}}
\def\Lfg{\mathfrak{f}}
\def\mO{\mathcal{O}}

\newcommand{\com}{\textcolor{red}}
\newcommand{\new}[1]{{\color[rgb]{1.0,0.,0}#1}}
\newcommand{\old}[1]{{\color[rgb]{0.7,0,0.7}\sout{#1}}}

\renewcommand{\real}{\ensuremath{\mathbb{R}}}

\newcommand*{\Cdot}[1][1.25]{%
  \mathpalette{\CdotAux{#1}}\cdot%
}
\newdimen\CdotAxis
\newcommand*{\CdotAux}[3]{%
    {%
	\settoheight\CdotAxis{$#2\vcenter{}$}%
	\sbox0{%
	    \raisebox\CdotAxis{%
		\scalebox{#1}{%
		    \raisebox{-\CdotAxis}{%
			$\mathsurround=0pt #2#3$%
		    }%
		}%
	    }%
	}%
	\dp0=0pt %
	\sbox2{$#2\bullet$}%
	\ifdim\ht2<\ht0 %
	\ht0=\ht2 %
	\fi
	\sbox2{$\mathsurround=0pt #2#3$}%
	\hbox to \wd2{\hss\usebox{0}\hss}%
    }%
}

\newcommand\hcancel[2][black]{\setbox0=\hbox{$#2$}%
\rlap{\raisebox{.45\ht0}{\textcolor{#1}{\rule{\wd0}{1pt}}}}#2} 

\renewcommand{\arraystretch}{2.5}%
\renewcommand{\floatpagefraction}{.8}%

\def\newthing{\marginpar{{\color{red}****}}}
\def\edit#1{{\color{red}{***#1***}}}
\reversemarginpar
\def\mN{{\mathbb N}}


\def\tu{\tau}
\def\ze{z}
\def\d{\partial}
\def\L{\varphi}  

\DeclareRobustCommand{\rchi}{{\mathpalette\irchi\relax}}
\newcommand{\irchi}[2]{\raisebox{\depth}{$#1\chi$}}


\hypersetup{pageanchor=false}
\begin{titlepage}
    \begin{flushright}	ICTS/2018/02, TIFR/TH/18-51
    \end{flushright}

    \vspace{.4cm}
    \begin{center}
      \noindent{\Large \bf{Gravitational collapse in SYK models and
          Choptuik-like phenomenon}}\\
      \vspace{1cm} Avinash Dhar$^{a}$, Adwait Gaikwad$^{b}$,
      Lata Kh Joshi$^{b}$, 
         Gautam Mandal$^{b,c}$, and Spenta
        R. Wadia$^{a,c}$

	\vspace{.5cm}
	\begin{center}
	    {\it a. International Centre for Theoretical Sciences}\\
	    {\it Tata Institute of Fundamental Research, Shivakote,
	    Bengaluru 560089, India.}\\
	  {\it b. Department of Theoretical Physics}\\
	    {\it Tata Institute of Fundamental Research, Mumbai 400005, 
	    India.}\\
	  {\it c. Kavli Institute for Theoretical Physics}\\ 
	  {\it University of California at Santa Barbara, CA 93106, USA}
	    \vspace{.5cm}
	\end{center}

	\gap2

	\today

    \end{center}

     \begin{abstract}

SYK model is a quantum mechanical model of fermions which is solvable at strong coupling and plays an important role as perhaps the simplest holographic model of quantum gravity and black holes. The present work considers a deformed SYK model and a sudden quantum quench in the deformation parameter. The system, as in the undeformed case, permits a low energy description in terms of pseudo Nambu Goldstone modes. The bulk dual of such a system represents a gravitational collapse, which is characterized by a bulk matter stress tensor whose value near the boundary shows a sudden jump at the time of the quench. The resulting gravitational collapse forms a black hole only if the deformation parameter $\Delta\epsilon$ exceeds a certain critical value $\Delta\epsilon_c$ and forms a horizonless geometry otherwise. In case a black hole does form, the resulting Hawking temperature is given by a fractional power $T_{bh} \propto (\Delta\epsilon - \Delta\epsilon_c)^{1/2}$, which is reminiscent of the `Choptuik phenomenon' of critical gravitational collapse. 

\end{abstract}

\vfill

\hrule
\gap{-3}
{\footnotesize     
\begin{flushleft}     
      {adhar@icts.res.in},
      {adwait@theory.tifr.res.in}, 
      {latakj@theory.tifr.res.in},
      {mandal@theory.tifr.res.in}, {spenta.wadia@icts.res.in}
      \end{flushleft}}
\end{titlepage}

\pagenumbering{roman}
\tableofcontents
\pagenumbering{arabic}
\setcounter{page}{1}

\newpage
\section{Introduction and Summary}\label{sec:intro}
The black hole information loss problem is usually associated with
black hole evaporation. However, even the process of formation of a
black hole can be regarded as a version of information loss. This is
because even if the collapsing matter is in a pure state, when it
forms a black hole it has an entropy. Hence a pure state appears to
evolve to a mixed (thermal) state; furthermore, the information about
the initial pure state appears to be lost. How does one understand
this puzzle within a unitary quantum mechanical framework?  With the
AdS/CFT correspondence, such a unitary description appears possible in
terms of the dual CFT where gravitational collapse can be modelled by
a quantum quench \cite{Chesler:2008hg,
  Bhattacharyya:2009uu,Hartman:2013qma,Anous:2016kss} and under a
sudden perturbation a given pure state can evolve to a pure state with
thermal properties \footnote{We will henceforth call such pure states
  with thermal properties, {\it thermal microstates}.}.  Such models
are not easy to construct in strongly coupled field theories in three
and higher dimensions.  In lower dimensions, however, there are
powerful techniques to deal with the dynamics of strongly coupled
conformal field theories. In one dimension, the relevant strongly
coupled model \cite{Sachdev:1992fk, Sachdev:2010um, Kitaev-talks:2015,
  Kitaev:2017awl, Engelsoy:2016xyb, PhysRevX.5.041025} which has a
holographic dual \cite{Almheiri:2014cka, Maldacena:2016upp,
  Jensen:2016pah, Mandal:2017thl} is the SYK model \footnote{The SYK
  model for charged fermions has been discussed in
  \cite{Davison:2016ngz} and the holographic dual to such a model has
  been presented in \cite{Gaikwad:2018dfc, sandip2}}.  In the present
paper, we will discuss gravitational collapse in such a holographic dual.

Besides the above issue of `information loss', gravitational collapse
is associated with another interesting phenomenon, namely that of
Choptuik scaling. In his classic work \cite{PhysRevLett.70.9} Choptuik 
analyzed a family of initial states characterized by a
parameter $p$ (which roughly corresponds to the amount of
self-gravitation of the infalling matter) and evolved them numerically
(see, e.g. \cite{Gundlach2007} for a review). He found that while no
black holes are formed for $p<p_c$, they {\it are} formed for $p>p_c$,
with the mass of the resulting black hole given by
$M_{bh}\propto(p-p_c)^\gamma$. Here, $\gamma$ is found to be a universal
critical exponent, which depends only on the type of infalling matter
and not on the details of the initial configuration. The results of
\cite{PhysRevLett.70.9} were in asymptotically flat space (for a
review see, e.g. \cite{Gundlach:1999cu}). This was generalized to
asymptotically AdS$_3$ spaces (a) for scalar field collapse in
\cite{Pretorius:2000yu} and (b)  for
formation of BTZ black holes in \cite{Birmingham:1999yt} from point particle collisions (the
critical exponents are different in the two cases).  In higher AdS$_D$
spaces, with $D\ge 4$, the Choptuik phenomenon gets richer, because
bounces from the boundary play a dominant role
\cite{Bizon:2011gg}\footnote{The critical value $p_c$ in this case
  refers to critical value of $p$ for black hole formation in the
  first pass for the imploding matter. For $p<p_c$ there are a new
  series of critical values $p_{c,n}$ so that for $p>p_{c,n}$ a black
  hole just about forms after $n$ bounces; the mass of the resulting
  black hole after $n$ bounces has the same critical behaviour $M_n
  \propto (p-p_{c,n})^\gamma$ with the same critical exponent
  \cite{Bizon:2011gg}. We thank M. Rangamani for important discussions
  on this issue.}.  It is important to obtain a field
theory understanding of Choptuik
criticality. SYK model allows us to explicitly perform a boundary
calculation dual to Choptuik transition, as we will describe in the
paper \footnote{Strictly speaking, we should call the phenomenon
  Choptuik-like since the precise details of a critical gravitational collapse
  are not yet possible to work out (see Section \ref{sec-collapse}
  for the extent to which details of the gravitational
  collapse are possible to work out)\label{ftnt-choptuik}.
  In the following, phrases like Choptuik phenomenon and Choptuik
transition will be used with this caveat in mind.} .

The basic framework for our study was laid out by Kourkoulou and
Maldacena in \cite{Kourkoulou:2017zaj}, where they described an
explicit construction of a complete set of thermal microstates for the
SYK model. The dual spacetime for such a state, evolving under the SYK
Hamiltonian $H_{SYK}$, is given by a black hole in the Poincare
wedge of AdS$_2$ with a special boundary condition at the corner. It
was shown in \cite{Kourkoulou:2017zaj} that, if one starts with a
given such state and turns on a certain perturbation $\ep H_M^{(s)}$ with a
suitably fine-tuned `spin's and coefficient $\ep$ higher than a
certain value (we will refer to this as the {\it first quench}), then
the state turns non-thermal and the horizon disappears\footnote{This
  perturbation is an example of a state-dependent operator which has
  been used in the context of understanding microstates in the black hole
  interior in \cite{Papadodimas:2013jku}; see also \cite{Maldacena:2017axo, Krishnan:2017txw, Goel:2018ubv, Almheiri:2018xdw}.} \footnote{If the coefficient $\ep$ is lower than the threshold we get a smaller black hole than the black hole that would have formed with $\ep=0$, see \cite{Brustein:2018fkr}. }.
  This is a {\it
  black hole $\to$ AdS transition}, which {\it prima facie} appears to
go against the second law of black hole thermodynamics! It turns out,
however, that the perturbation Hamiltonian contributes {\it negative
  energy}, thereby allowing reduction of black hole entropy.

The objective of our paper is to start with the above mentioned
non-thermal state of the system (corresponding to AdS geometry) and
perform a \textit{second quench}, in which we decrease the coefficient $\ep$ of
the perturbation $H_M$ by an amount $\Delta \ep$. The calculation is
done in the boundary theory, which can be interpreted in terms of the
dynamics of the horizon. We find that there is a critical value
$\Delta\epsilon_c$ of the coefficient $\Delta\epsilon$, beyond which
the AdS geometry turns into a black hole. Decreasing $\ep$
i.e. having $\Delta \ep>0$ contributes positive energy to the system and the
above process can be identified with gravitational collapse of matter.
The existence of the critical value can be identified with a Choptuik
phenomenon for gravitational collapse in AdS$_2$ \footnote{In two dimensions,
in the context of two-dimensional dilaton-gravity black holes
    \cite{Mandal:1991tz,Witten:1991yr,Elitzur:1991cb},
    gravitational collapse has been discussed in
    \cite{Strominger:1993tt}.} We find that for
$\Delta\epsilon > \Delta\epsilon_c$ the temperature of the resulting
black hole is given by $T_{BH} \propto (\Delta\epsilon -
\Delta\epsilon_c)^{\f12}$, corresponding to an exponent $\f12$. 
\footnote{Preliminary results were presented at \url{https://indico.cern.ch/event/691363/timetable/\#14-gravitational-collapse-in-t}},\footnote{Quantum quenches in SYK model have been discussed, in somewhat different contexts, in \cite{Eberlein:2017wah} and \cite{Bhattacharya:2018fkq}.} 

\paragraph{Outline and summary of results}

\begin{enumerate}
\item {In section \ref{sec-review} we review the pure states in the SYK model as presented in the Ref., \cite{Kourkoulou:2017zaj} .}

\item The deformed SYK model is described by an effective action, involving a modified Schwarzian \eq{full-action} which was presented in Ref. \cite{Kourkoulou:2017zaj}. We include a detailed path integral derivation of this action in appendix \ref{app-path-integral}

\item {We use the above effective action \eq{full-action}  to construct solutions of the equation of motion in the deformed SYK model.} As mentioned in the introduction, the thermal microstate in the bulk describes a bulk with a black hole when evolved with the SYK Hamiltonian $(\ep =0)$. In order to get a Poincare patch without a black hole in the bulk, the Kourkoulou-Maldacena perturbation should be fine tuned with respect to the thermal microstate. (Section \ref{fine-tuning})    

\item
  We construct the Hamiltonian corresponding to the modified
  Schwarzian.  This follows from a Noether procedure for time
  translation symmetry. We show that for positive coefficient of the
  perturbation, the Hamiltonian has a binding energy i.e. the
  classical vacuum has a negative energy $-m_{gap}$ (see Section
  \ref{sec-mass-gap}). To reach positive energy, one has to supply
  at least this amount of energy, i.e. decrease $\ep$ by more than
  a critical amount.

\item
  We provide a bulk description of the above phenomenon in section \ref{sec-bulk}. The bulk
  geometry corresponding to the above classical vacuum is horizonless
  and is AdS$_2$-Poincare \footnote{With a special boundary condition
    at the corner of the Poincare wedge.} . The energy is negative
  (viz. $-m_{gap}$) because of contributions of matter stress tensor.
  In Section \ref{sec-collapse} we show that performing a sudden
  quench in the boundary theory corresponds to sudden release of bulk
  matter from the boundary. From the expression for the ADM mass
  \eq{h-f}, it is clear that decreasing $\ep$ increases ADM mass of
  the system. Combining with the above discussion of bulk matter, a
  decrease of bulk $\ep$ therefore corresponds to release of positive
  energy matter from the boundary. In order to create a black hole,
  one has to reach positive ADM mass, and hence must supply a minimum
  energy (at least $m_{gap}$); this translates to the fact that the
  change of $\ep$, $\Delta\ep$, must exceed a critical value (denoted by $\Delta \ep_{cr}$). It is
  shown in Section \ref{sec-choptuik} that the effective temperature
  goes as $T_{bh} \propto \sqrt{\Delta\ep - \Delta\ep_{cr}}$. 

\item
  The results in Section \ref{sec-bulk}  rely on boundary dynamics through
  two quantum quenches. These calculations are done in Section
  \ref{app-onesidsol}. As in \cite{Kourkoulou:2017zaj}, a potential
  picture is used to simplify the dynamics. The bottom of the
  potential corresponds to the classical vacuum and the mass gap
  mentioned above corresponds to the depth of the bottom.

\item
  The main dynamical quantity is the pseudo-Nambu Goldsone mode $f(t)$  which has a direct meaning in the bulk. Besides computing $f(t)$, we  derive some correlation functions as they evolve dynamically. The change to the black hole geometry in the bulk corresponds to a
  change from power law behaviour to exponential decay. We identify
  this with thermalization, which is the characteristic of black
  holes. These calculations are presented in Section \ref{sec-2pt}.

\item In the final Section \ref{mal-qi}, we show that the Choptuik
  phenomenon occurs also in the two-sided model \cite{Maldacena:2018lmt} in the context of gravitational collapse from global AdS$_2$ to eternal black hole.

  \end{enumerate}

\section{Pure states of the SYK model}\label{sec-review}

In this section we briefly review the relevant properties of the pure states of \cite{Kourkoulou:2017zaj}.
We will start with the SYK Hamiltonian which is written in terms
of $N$ Majorana fermions
\begin{equation}\label{eq:ham0}
  \text{H}_0 = \text{H}_{SYK} =-\sum_{1\le a<b<c<d \le N} j_{abcd}\ \psi_a\psi_b\psi_c\psi_d
\end{equation}
where the couplings $j_{abcd}$ are drawn randomly from a Gaussian distribution with zero mean and variance, $\braket{J_{abcd}} = 3! J^2/ N^3$. 
The equal time anticommutation relation of the Majorana fermions, $\{
\psi_a, \psi_b\}= \delta_{ab}$, coincides with the $SO(N)$ Clifford
algebra. We will call the normalized states which provide a spinorial
representation of the above algebra, $|B_s\ran$, where $s= (\pm, \pm,
...)$ are $N/2$ dimensional `spin' vectors. The total number of such
vectors is $2^{N/2}$.

Ref. \cite{Kourkoulou:2017zaj} introduced a class of pure states (similar
to the Calabrese-Cardy states \cite{Calabrese:2005in} that were
introduced to model quantum quench) given by
\[| B_s(l)\ran = \exp[-l H_0]|B_s\ran,\]
which reproduce thermal properties for a large class of
observables, corresponding to a temperature $T=1/\b$ {\it where
  $\beta=2l$}.  E.g.
\begin{align}
  &\lan B_s(l)| B_s(l) \ran = \lan B_s|e^{-2l H_0}|B_s \ran = 2^{-N/2}\Tr(e^{-\b H_0})
  \equiv 2^{-N/2}Z(\beta)
\label{thermal-a}
\\
&\quad \lan B_s(l)|\psi_a(t) \psi_a(t')| B_s(l) \ran =
G_\b(t-t') \equiv \f{C_\Delta}{\left[\f{J\be}{\pi}\sinh(\pi(t-t')/\b)\right]^{2\Delta}}
\label{thermal-b}
\end{align}
Eqs \eq{thermal-a}, \eq{thermal-b} are obtained at large $N$, where the fermion bilinear and $H_0$ are invariant under the flip group which changes the spin vector $s$ (See Appendix \ref{app-path-integral}). These equations
show that the basic dynamical variable of the SYK model, the bilocal variable $G(t,t')= (1/N)\sum_a \psi_a(t) \psi_a(t')$ which describes the $O(N)$ invariant sector, does not distinguish between the pure states $|B_s(l)\ran$ and the thermal (mixed) state $\rho_\b = \f1{Z(\b)} \exp[-\b H_0]$, $\b=2l$.

Equalities like \eq{thermal-a} can be obtained from a path-integral \cite{Kourkoulou:2017zaj} (a detailed derivation is given in Appendix \ref{app-path-integral}). For large enough $l$ ($lJ\gg1$), both sides of the equation \eq{thermal-a} can be expressed as follows in terms of a path-integral over the time-reparameterization mode $f(\tau):= \f\pi{\b
  J^2} \tan(\pi \varphi(\tau)/\b)$
\begin{align}
  \int [Df] \exp[i S_0[f]], \quad S_0[f]= -\f{N \alpha_s}{J}
  \int_{-l}^{l} d\tau \, \{f,\tau \} = -\f{N \al_s}{J}
  \int_{-l}^{l} d\tau \,[ \{\varphi(\tau),\tau\} + 2\f{\pi^2}{\beta^2}
  (\varphi'(\tau))^2].
\label{path-integral-Z}
\end{align}
In the above we have used the notation for the Schwarzian of a function $f(\tau)$
\[
\{ f, \tau \} \equiv \f{f'''}{f'}- \f32 \left(\f{f''}{f'}\right)^{\!2},
\]
where prime denotes derivative with respect to $\tau$. The boundary condition for the path integral to describe the LHS of \eq{thermal-a} is appropriate for an interval (see \cite{Kourkoulou:2017zaj}):
\begin{align}
&\varphi(-l)=-l,
  \varphi(l)=l,  \varphi'(-l)= \varphi'(l)=1.
  \label{interval-bc}
  \end{align}
The boundary condition for the circle is given by periodic
identification of $\tau=-l$ and $\tau=l$ and winding number 1.  The
saddle point solutions for the two boundary conditions are different;
however, the classical action $S_0[f]$ evaluates to the same value in
both cases (the two boundary conditions differ in the SL(2) zero
modes, which are gauge modes and do not affect physical
quantities). Both boundary conditions also lead to the same result for
Green's functions e.g.  \eq{thermal-b} or \eq{off-diagonal}, which
explains why $|B_s(l)\ran$ states reproduce thermal properties despite
different boundary conditions.\footnote{We acknowledge crucial
  discussions with Juan Maldacena regarding the issue of boundary
  conditions.}
\par The saddle point solution corresponding to the boundary condition \eq{interval-bc} is $\varphi(\tau) = \tau$. In terms of the $f$ variable the solution is 
\begin{equation}\label{saddle-interval}
 f(\tau) = \f{\pi}{\b J^2}\ \tan\left({\f{\pi}{\b}\ \tau}\right).
\end{equation}
Eq.\eq{thermal-b} demonstrates diagonal observables which show thermal properties in the $|B_s(l)\ran$ states.  What about off-diagonal bilinears?  An important relation is
\begin{align}
\lan B_s(l)| s_k \psi_{2k-1}(t) \psi_{2k}(t') |B_s(l) \ran =-2 i G_\b(t-il)
G_\b(t'-il),
\label{off-diagonal}
\end{align}
which shows that the off-diagonal bilnears $\psi_{2k-1}(t) \psi_{2k}(t')$ by themselves do {\it not} have a thermal form (and depend on the spin vector $s$); but when they are in the combination as shown above ($s_k \psi_{2k-1}(t) \psi_{2k}(t')$), then the RHS is thermal and the memory of the spin vector $s$ is erased in the RHS.  A related property which can be derived using similar methods as above is
\begin{align}
\lan B_{s'}(l)| s_k \psi_{2k-1}(t) \psi_{2k}(t')| B_{s'}(l) \ran =-2i \cos\theta\, G_\b(t-il) G_\b(t'-il),\quad  \; s.s'= N/2\ \cos\theta
\label{cos-alpha}
\end{align}
This shows that unlike in \eq{off-diagonal}, if the spin vectors in the state and operator are not matched, some memory of both spin vectors is retained.

\section{Modified SYK model and the effective action}\label{sec-perturbation}
We now consider deforming the SYK theory with a bilinear operator
\cite{Kourkoulou:2017zaj}
\begin{equation}\label{eq:ham}
H = {H}_0 + \ep(t) {H}^{(s)}_M, \quad {H}_0 =
{H}_{\text{SYK}}, \quad {H}^{({s'})}_M = - i J \sum_{k=1}^{N/2}\ s'_k \psi_{2k-1} \psi_{2k} = - \left(\f{J}{2}\right) \sum_{k=1}^{N/2}\ s'_k \hat{S}_k .
\end{equation}
In the previous section, we considered the path integral \eq{thermal-a} which could be regarded as a Euclidean time evolution between a pair of $|B_s\ran$ states. Let us now consider a path integral containing a real time evolution, with $\ep(t)$ as above, between a pair of $|B_s\ran$ states. Such an evolution can be regarded as over a 'closed time path' in the complex time plane ${z}=t+i\tau$ Fig\ref{ctpmain}.
\begin{figure}[H]
 \centering
 \begin{tikzpicture}
  \draw [->](3,-2.5)--(3,3);
  \draw [->,red,thick] (3,2)--(3,1);
  \draw [red,thick](3,1)--(3,0.5);
  \draw [->,red,thick] (3,-.5)--(3,-1);
  \draw [red,thick] (3,-1)--(3,-2);
  \draw [->](0,0)--(13,0);
  \draw [->,thick](3,.5)--(6,.5);
  \draw [thick](3,.5)--(12.5,.5);
  \draw [->,thick](12.5,-.5)--(6,-.5);
  \draw [thick](6,-.5)--(3,-.5);
  \draw[fill,red,thick] (3,2) circle [radius=0.08];
  \draw[fill,red,thick] (3,-2) circle [radius=0.08];
  \node [below right] at (3,0) {$ t =0$}; 
  \node [below right] at (11,0) {$t \to \infty$};
  \draw[blue] (12.5,.5) to [out=300,in=60] (12.5,-.5);
  \draw[blue,very thick,dashed] (0,2)--(10,2);
  \draw[blue,very thick,dashed] (0,-2)--(10,-2);
  \draw[decoration={brace,raise=5pt},decorate] (3,.5) -- node[left=8pt] {$l$} (3,2);
  \draw[decoration={brace,mirror,raise=5pt},decorate] (3,-.5) -- node[left=8pt] {$l$} (3,-2);
 \end{tikzpicture} 
\caption{ The red dots are the states $\ket{B_s}$ (above) and $\bra{B_s}$ (below). If we have insertions of operators which are flip symmetric then the boundary condition (red dots) can be converted to trace boundary condition. In such a case we identify the two blue lines and the Euclidean path of contour becomes a circle.}
\label{ctpmain}
\end{figure}
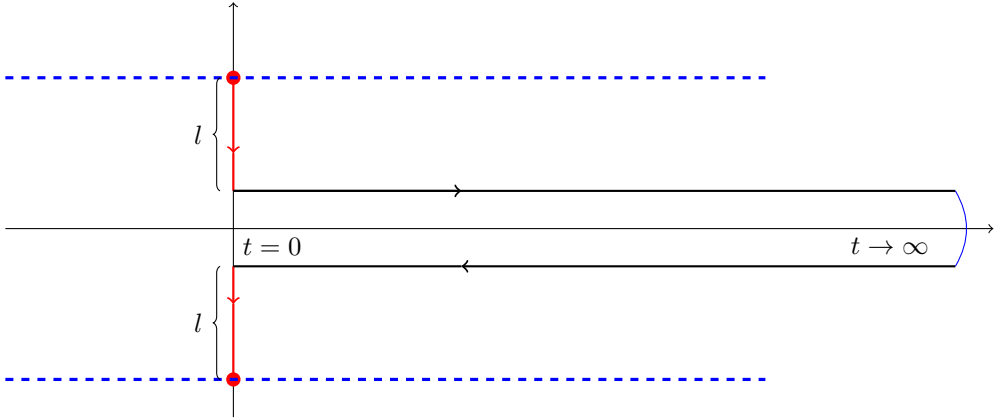
The path integral in the large $N$ limit along the real time part of the contour is governed by \cite{Kourkoulou:2017zaj} (see Appendix \ref{app-path-integral} for details)
\begin{align}
  S[f]= -\f{N \alpha_s }{J}\int dt\
  \left[\{f(t),t\} -   \f{J^2\hat{\epsilon}(t)}{2} \left(f'\right)^{2\Delta}\right].
\label{full-action}
\end{align}
In the present work we work with, $\Delta=1/4$. The part of the contour which is Euclidean is still governed by action given in \eq{path-integral-Z} (see Appendix \ref{app-path-integral} for details). We aim to solve the classical dynamics described by $S[f]$. The boundary conditions for the real time part are obtained from the classical solution on the Euclidean part. We have also defined a convenient variable $\hat{\e}(t) = (2 C^2_\Delta\ \cos\theta/\alpha_s) \e(t)$, where $\cos\theta$ is defined in \eq{cos-alpha} and time dependence of $\hat{\ep}(t)$ is as follows (see Figure \ref{fig-quench-protocol})
\begin{equation}\label{ehat}
 \hat{\epsilon}(t) = \epsilon_1 \Theta(t) + (\epsilon_2-\epsilon_1) \Theta(t-T).
\end{equation}
\begin{figure}[H]
 \centering
 \begin{tikzpicture}
  \draw [->] (0,-1)--(0,3);
  \node [below right] at (0,0) {$t=0$};
  \node [left] at (0,2.5) {$\hat{\ep}(t)$};
  \draw [->] (-1,0)--(9,0);
  \node [below] at (10,0) {$t$};
  \draw [dashed,red, very thick] (-1,.8)--(1,.8);
  \draw [dashed,red, very thick] (3.5,1.3)--(5.5,1.3);
  \node [left] at (-1,.8) {$\ep^0_{\text{cr}}$};
  \node [right] at (5.5,1.3) {$\ep_{\text{cr}}$};
  \draw [blue,very thick] (0,0)--(0,2);
  \draw [blue,very thick] (0,2)--(4.5,2);
  \node [above] at (2,2) {$\hat{\ep}=\ep_1$};
  \draw [blue,very thick] (4.5,2)--(4.5,.5);
  \draw [dashed](4.5,3)--(4.5,-1);
  \node [below right] at (4.5,0) {$t=T$};
  \draw [blue,very thick] (4.5,.5)--(8.5,.5);
  \node [above] at (6,.5) {$\hat{\ep}=\ep_2$};
  \draw [->,very thick] (6,2.5)--(4.6,1.4);
  \node[right] at (6,2.5) {Choptuik transition};
  \end{tikzpicture}
\caption{Quench protocol used in this paper. A non-zero value of
  $\hat{\ep}=\ep_1$ is turned on, as in \cite{Kourkoulou:2017zaj}, at time
  $t=0$ (which we call the first quench). After a time $t=T$, a second
  quench is performed, to $\hat{\ep}=\ep_2$.}
\label{fig-quench-protocol}
\end{figure}
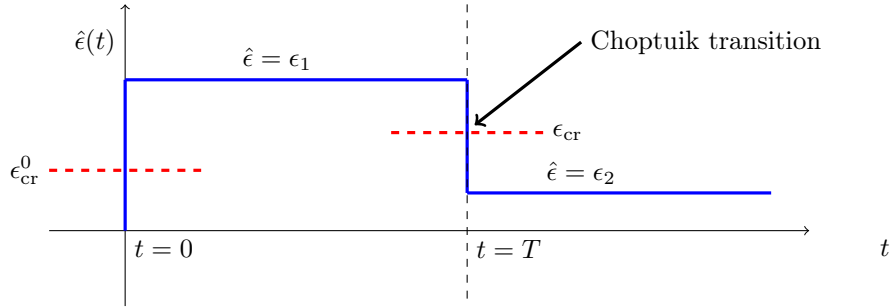
\subsection{Fine tuning}\label{fine-tuning}
In this subsection, we will show that in order to keep $\hat\ep(t)\sim O(1)$, we have to fine tune spin vector $s'$ that appears in $H^{(s')}_M$ w.r.t. the spin vector $s$ that appears in $\ket{B_s}$ to an accuracy of $O(e^{-N})$. To see this, note that $\hat\ep(t)$ is proportional to $\cos\theta = (2/N)s.s'$. Now if we assume that $r$ spins are different between $s'$ and $s$ then
\begin{equation}
 \cos\theta = 1-\f{4r}{N}.
\end{equation}
 The number of states for which $s$ and $s'$ differ by $r$ spins is $^{N/2}C_r$ and the fraction of such states is given by $P(r) = 2^{-N/2}\ (^{N/2}C_r)$. We can approximate $P(r)$ as 
\begin{equation}\label{P(r)}
 P(r) = 2^{-N/2}\ \left(^{N/2}C_r\right) \approx \s{\f{1}{2 \pi \sigma^2}}\ \exp\left[{-\f{1}{2\sigma^2}\ \left(r-\braket{r}\right)^2}\right] , \quad \braket{r} = \f{N}{4} \text{ and } \sigma = \s{\f{N}{8}}. 
\end{equation}
To arrive at \eq{P(r)} we used Stirling's approximation for the binomial coefficient. At large $N$, $r$ effectively becomes a real continuous variable ($-\infty < r < \infty$). 
\begin{figure}[H]
 \centering
 \begin{subfigure}[b]{0.4\textwidth}
\centering
 \includegraphics[width=\linewidth]{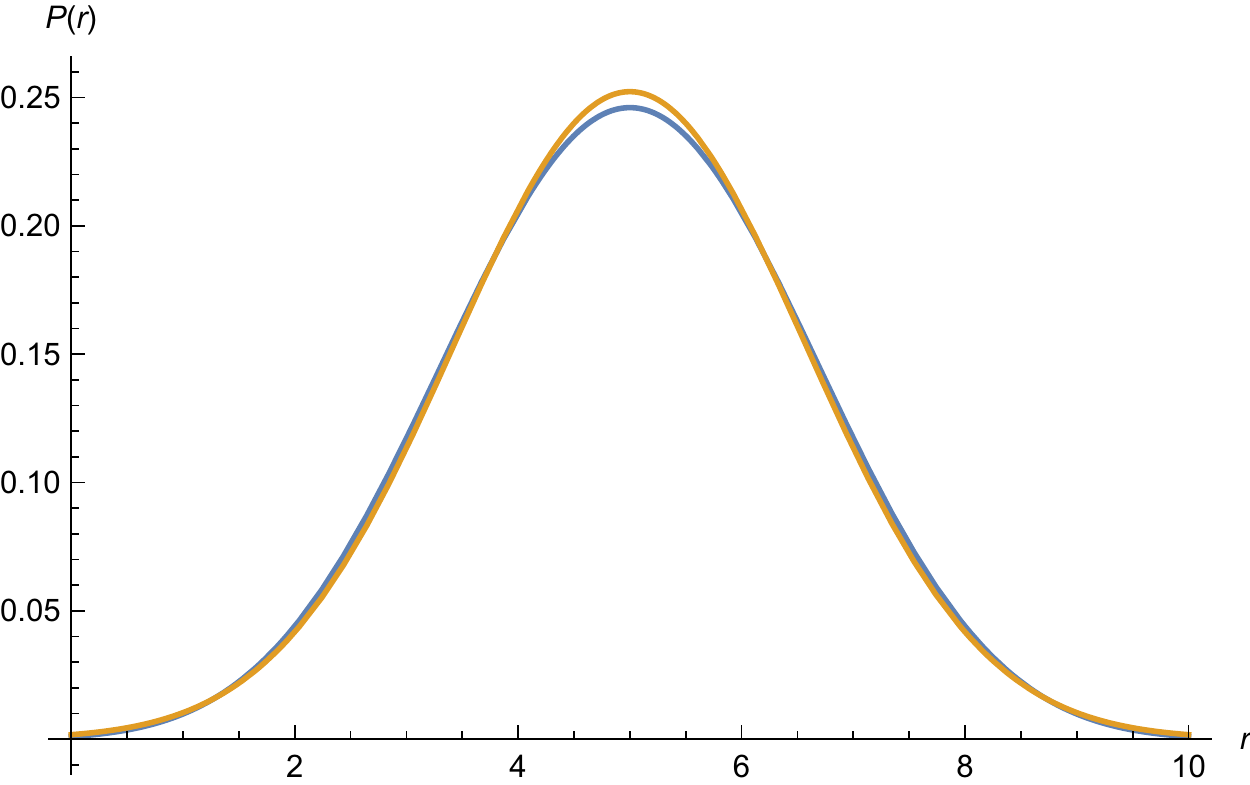}
\caption{$N=20$}
\end{subfigure}
\hspace{2cm}
\begin{subfigure}[b]{0.4\textwidth}
\centering
 \includegraphics[width=\linewidth]{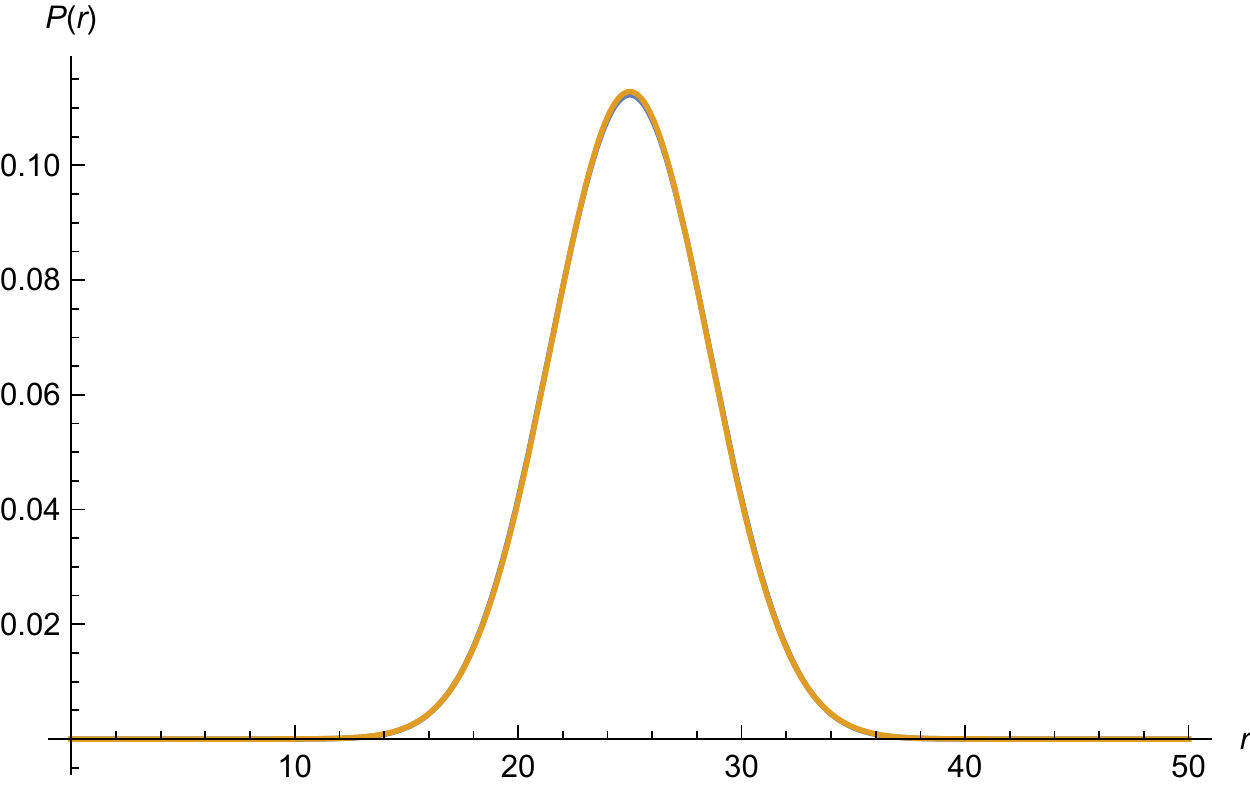}
 \caption{$N=100$}
\end{subfigure}
 \caption{The plot of exact (blue) and approximate (orange) $P(r)$ for a) $N=20$ and b) $N=100$ (the curves are overlapping). At large $N$ distribution is sharply peaked around the average and the approximation is more accurate.}
\end{figure}
The value of $\cos\theta$ decreases as $r$ increases, and it remains positive up to $r=N/4$ when $\cos\theta=0$ and when $r>(N/4)$ then $\cos\theta<0$. We want $\cos\theta \sim O(1)$ and positive, at large $N$ the distribution is sharply peaked around the average ($r=\braket{r}$) hence we could go very close to $\braket{r}$ e.g. $r = (N/4) - 3\sigma$ and still wound not have a significant fraction of states $P(r)$. The total fraction of states $(I)$ for which $\cos\theta > \cos\theta_m$ is given by 
\begin{equation}\label{I}
 I = \int_0^{r_m} P(r) dr = \f{1}{2}\left(1-\text{Erf}\left(\f{\s{N}\ \cos\theta_m}{2}\right)\right), \quad r_m = \f{N}{4}(1-\cos\theta_m).   
\end{equation}
Here Erf$(x)$ is error function. Now we will take large $N$ limit assuming $\cos\theta_m \sim O(1)$ which gives us 
\begin{equation}\label{Iexp}
 I = \f{e^{-N \cos^2\theta_m/4}}{\pi \s{N} \cos\theta_m} + O\left(N^{-3/2}\right).
\end{equation}
This tells us as that fraction of states for which $\cos\theta \sim O(1)$ at large $N$ is exponentially small in $N$ and in order to arrange $\hat \ep(t) \sim O(1)$ we have to choose $s'$ from this exponentially small fraction of total number of possible states $s'$. Note that if we consider $\cos\theta_m \sim N^{-p}$ and when $p < 1/2$ then we have result \eq{Iexp}, but when $p=1/2$ then $I \sim O(1)$ and when $p>1/2$ then $I \to 1/2$.

\subsection{Equations of motion}
We can now look at the equation of motion for the modified action $S[f]$ is
\begin{equation}\label{mseom}
 \{f(t),t\}' = J^2 (f'(t))^{2\Delta} \Delta \left[\hat{\epsilon}'(t) + \hat{\epsilon}(t) (2\Delta -1) \left(\f{f''(t)}{f'(t)}\right) \right].
\end{equation}
Since $\hat{\ep}(t)$ is piecewise constant, the solution of \eq{mseom} can
be obtained in a piecewise fashion along the complex contour
${{\mathcal C}}$ in Fig \ref{ctpmain}, with appropriate
matching conditions across segments. Although it is possible to find
these solutions directly in terms of $f(z)$, it is more
instructive to write the above path integral \eq{full-action} in terms
of a variable $\phi:= \ln(f')$ ($f'$ is non-negative), leading to a new action principle
which is second order in time derivatives \cite{Kourkoulou:2017zaj}:
\begin{equation}\label{mSeff}
 S = \frac{N \alpha_s}{J} \int dt \left[\frac{\phi'^2}{2} + \f{J
     \lambda(t)}{2} (e^{\phi} - f') + \f{J^2
     \hat{\epsilon}(t)}{2}\ e^{2\Delta \phi} \right].
\end{equation} 
Here $\lambda(t)$ is a Lagrange multiplier. The equations of motion
from \eq{mSeff} are
\begin{align}
\phi''(t) &=  J^2\hat{\epsilon}(t)\Delta e^{2\Delta \phi(t)} + \f{J \lambda(t)}{2} e^{\phi(t)} = - \f{dV(\phi)}{d\phi},\label{phieom}\\
\lambda'(t)&= 0, \quad
f'(t) =e^{\phi(t)}. \label{rest-eom}
\end{align}
These are, of course, equivalent to \eq{mseom}; in particular, both
sets of equations, \eq{mseom} and \eq{phieom}-\eq{rest-eom} clearly have four
integration constants. 

The potential $V(\phi)$ appearing in the equation of motion of $\phi$ is given by
\begin{align}
V(\phi) = \left(\f{J(- \lambda)}{2}\right) e^{\phi} - \left(\f{J^2
  \hat{\epsilon}(t)}{2}\right)\ e^{2 \Delta \phi}.
\label{v-phi}
\end{align}
In every segment of time where $\hat{\e}$ is a constant, we have a
constant value of the energy (recall that $\lambda$ is also a constant
by the equation of motion; in fact as we will see below, it will turn
out to be negative)\footnote{We have omitted $N\al_s/J$ from
  the expression for the `energy' $E$ to avoid clutter. To get the real
  energy we must multiply  the RHS of \eq{EC} by $N\al_s/J$.\label{ftnt-energy}}
\begin{equation}\label{EC}
 E \equiv \f{\phi'^2}{2} + V(\phi) = \f{\phi'^2}{2} + \f{J(- \lambda)}{2}\ e^{\phi(t)} - \f{J^2 \hat{\epsilon}}{2}\ e^{2 \Delta \phi(t)} .
\end{equation}
Depending on the sign of $\hat{\epsilon}$ two shapes of
potential are possible
 \begin{figure}[H]
 \centering
\begin{subfigure}[b]{0.4\textwidth}
\centering
 \includegraphics[width=\linewidth]{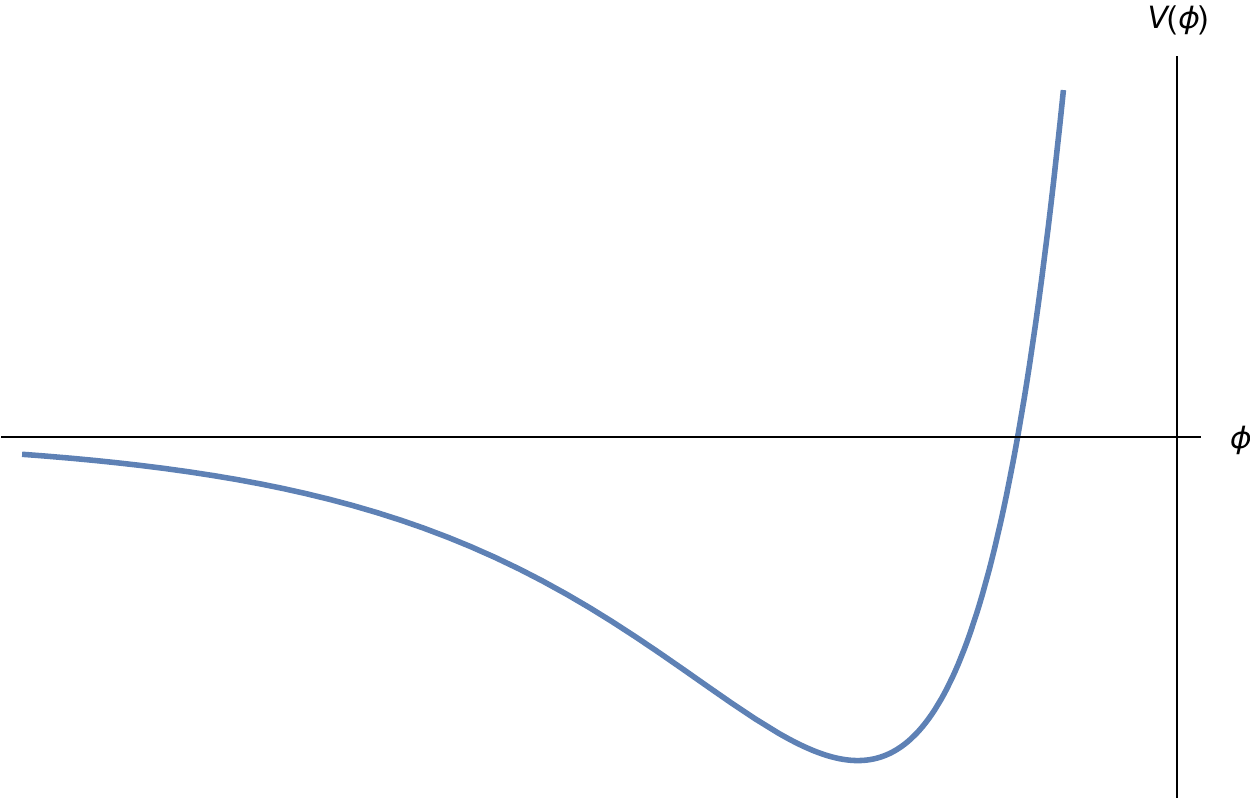}
\caption{$\hat{\epsilon} >0$}
\end{subfigure}
\hspace{2cm}
\begin{subfigure}[b]{0.4\textwidth}
\centering
 \includegraphics[width=\linewidth]{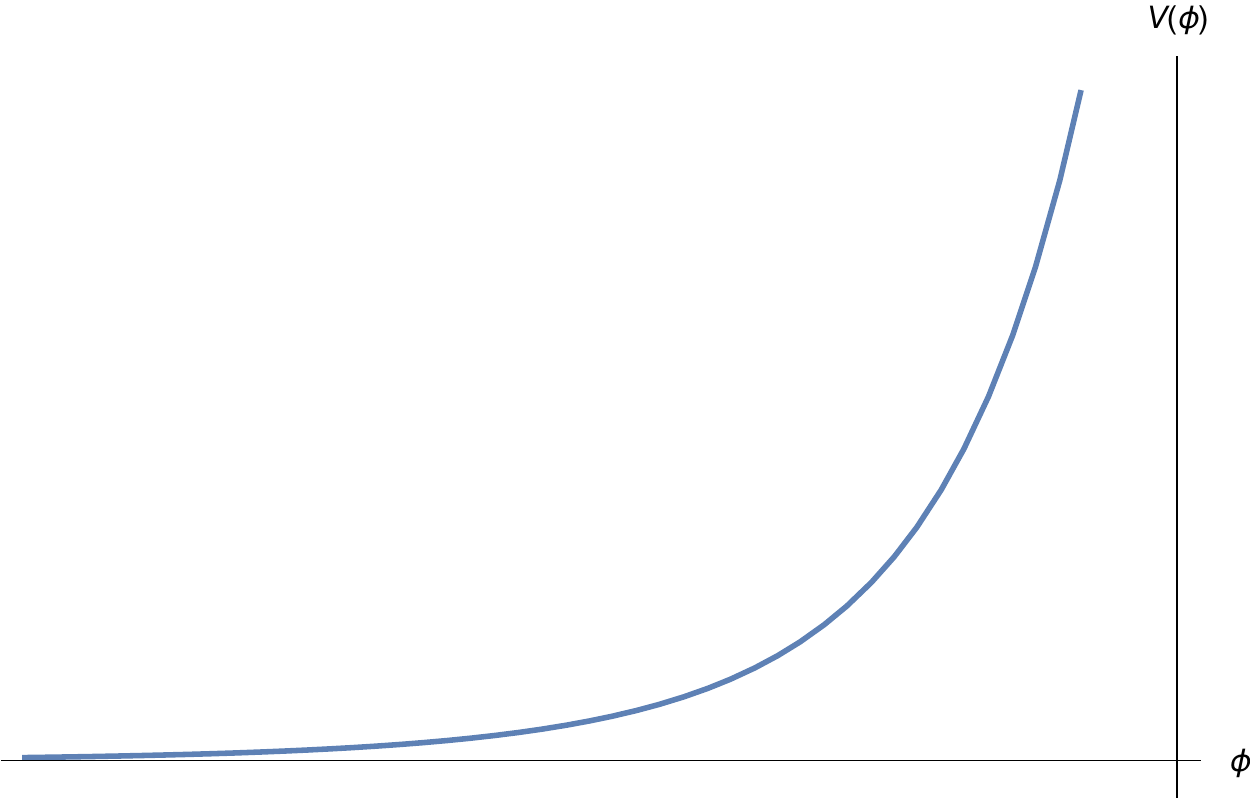}
 \caption{$\hat{\epsilon} \le 0$}
\end{subfigure}
\caption{Possible shapes of the potential, depending on the sign of
  $\hat \epsilon$. We have bounded or scattering solutions depending on whether
  $E<0$ or $E>0$. It is assumed for simplicity that
  $\Delta <\f12$. }
\label{potential}
\end{figure}
In each segment where $\hat\ep$ is constant, it is possible to directly
construct the Hamiltonian for the action \eq{full-action} as conserved
Noether charge for time translation. Such a Hamiltonian is
\begin{align}
H_f =\frac{N \alpha_s}{J}[ -\{f,t\} + \left(\Delta - \f12\right) \hat\epsilon J^2 {f'(t)}^{2\Delta}]
\label{h-f}
\end{align}
The value of this charge can be identified with that of the
ADM mass in the bulk. This Hamiltonian agrees onshell with \eqref{EC}.

\subsection{The mass gap}\label{sec-mass-gap}

It is easy to show that $H_f$ has a minimum value, given by the
minimum value of \eq{EC} or equivalently that of the potential
\eq{v-phi}, namely (where we have reinstated the factor $N\al_s/J$)
\begin{align}
  m_{gap}\equiv - H_{f,min}=- V_{min}= N \alpha_s J \hat\ep
  \left(\f12 -\Delta\right)
  \left( \f{2\hat{\ep}\Delta J}{-\lambda}\right)^{\f{2\Delta}{1-2\Delta}}
\label{h-min}  
  \end{align}
In order to reach $E>0$ from this minimum energy configuration, one
has to supply at least the amount of energy $m_{gap}$ (this is
analogous to ionization energy).

We will interpret this as the minimum energy, starting from the
classical vacuum of \eq{h-f}, to be able to form a black hole.
We will find that one way to supply energy to the system is
to decrease $\ep$ beyond a certain critical value; this will
be responsible for a Choptuik phenomenon in our context.

\section{Classical solutions}\label{app-onesidsol}

In this section, we will solve the classical equations of motion
\eq{phieom}-\eq{rest-eom}. As indicated above, since $\hat\epsilon(t)$ is
piecewise constant, we will solve for $\phi(t)$ in three segments: (a)
The Euclidean time segment $\tau \in [-l, 0]$\footnote{$\tau = -\text{Im}[z],$ where $z$ complex time.}, (b) The first real time
segment $t\in [0,T]$, and (c) The final real time segment $t>T$.  The
matching condition across the constant-$\hat\epsilon$ segments, as
dictated by the above equations of motion, is given by the continuity
of $\phi(t)$ and $\phi'(t)$, as well as that of $f$ and $\lambda$.

For the sake of simplicity, we will choose below $\Delta=\f14$,
although the essential features will remain true for any other value
of $\Delta < \f12$.

\paragraph{The Euclidean time segment}

In this segment $\hat \epsilon=0$; hence the solution $f=f_0(\tau)$ (and
consequently $\phi_0(\tau)= \ln(f_0'(\tau))$) are given by
\eq{saddle-interval}. Analytically continuing to ${\tt t}=t- i \tau$,
and projecting on to the negative real line ${\tt t}=t<0$, we get
\begin{align}\label{ebh}
 \phi_0(t) &= \log\left(\f{\pi^2}{\be^2 J^2} \sech^2\left(\f{\pi t}{2 l}\right)\right), \quad t<0.
\end{align} 
By the equations of motion, $\lambda(t)= \lambda_0$ is a
constant.  Eq. \eq{phieom} with
$\hat\epsilon=0$ gives $\lambda_0 =\f2{J} \phi_0''(t) e^{- \phi_0(t)}$. By
substituting the above solution for $\phi_0(t)$, we get
\begin{align}
  \lambda_0 = - 4 J.
  \label{lambda-value}
  \end{align}
By continuity across segments $\lambda$ retains this constant value
throughout. Henceforth we will put $\lambda=-4 J$ in \eq{phieom}.

\paragraph{General structure of the real time solutions}

Using $\lambda = -4J$ and $\Delta = 1/4$ in \eq{EC}, it is easy to get
\begin{align}\label{quad}
 -\int^{t}_{t_0} d\tilde t &= \int_{\phi_0}^{\phi} \frac{d\tilde \phi}{\sqrt{2 E - J^2(4 e^{\tilde \phi} - \hat{\epsilon} \ e^{\tilde \phi/2})}}\\
\hbox{hence,}~ -2J(t-{t_0}) &= \int_{\phi_0}^{\phi} \frac{d\tilde \phi}{\sqrt{(e^{\tilde \phi/2} - r^-)(r^+ - e^{\tilde \phi/2})}}\nonumber\\
 r^\pm &= \left(\f{\hat{\epsilon}}{8} \pm \f{\sqrt{32 E + J^2 \hat{\epsilon}^2}}{8\ J}\right) = \f{\hat{\epsilon}}{8} \left(1\pm \sqrt{\hat{E} + 1}\right) \quad \text{here }\hat{E} = \f{32 E}{J^2 \hat{\epsilon}^2}. \nonumber
 \end{align}
Here $r_\pm$ are the two roots for $e^{\tilde \phi/2}$ of the equation, $E= V(\tilde \phi)$. If both are positive, which happens for $E<0$, these represent two turning points (see Fig \ref{potential}). In this case the motion of $\phi(t)$, and consequently that of $f(t)$, is bounded and oscillatory. For $E>0$, only $r_+$ is positive ($r_-$ is negative and is unphysical)-- hence there is only one turning point, implying that $\phi(t)$ is only bounded above and is unbounded below. We will shortly discuss the interpretation of these cases in the bulk geometry.

\underline{Solving the integral:} 
Making the substitution $e^{-\phi/2} = x^2 + \frac{1}{r^+}$ the integral
can be reduced to a standard one. After some manipulations, we get
\begin{equation}\label{ephi}
 e^{\phi(t)} =  \left(\f{4 E}{J}\right)^2 \f{1}{\left(\sqrt{J^2 \hat{\epsilon}^2 + 32 E} \cos\left(\sqrt{\f{-E}{2}}\ (t-{t_0}) + 2\ \theta_0\right)-J\hat{\ep}\right)^{2}} = f'(t)
\end{equation}
where $\theta_0$ is defined by
\[
\sin^2\theta :=  \f{r_+ r_-}{r_+-r_-}\left(e^{-\phi_0/2}- \f1{r_+}\right).
\]
To write the solution for $f(t)$, let us define a set of new variables
\begin{align}\label{abcd}
 &a = \left(\f{4 E}{J}\right)^2, \quad b = \sqrt{J^2 \hat{\epsilon}^2 + 32 E}, \quad c =\sqrt{\f{-E}{2}}, \nn \\ 
 &d = 2\sin^{-1}\left(\s{\f{-2 E\ x^2_0}{J\ b}}\right), \quad x_0 = \s{e^{-\phi_0/2} - \f{8J}{J \hat{\ep} + b}}
\end{align}
with these definitions, $a$ and $b$ are always non-negative real numbers,
whereas $c,d$ are non-negative real for
$E<0$, and purely imaginary for $E>0$. This leads to two classes
of solutions depending on the sign of $E$:

$\bullet$ Case I: For $ E < 0$ (note that
$E> -\f{J^2 \hat{\ep}^2}{32} \equiv V_{min}$ which is the minimum of the
potential $V(\phi)$)
\begin{equation}\label{eq:fpp}
 f_p'(t) =  \frac{a}{\left(b \cos\left(c\ (t-{t_0}) + d\right)-J \hat{\ep}\right)^{2}},
\end{equation}
which gives us
\begin{equation}\label{fp}
 f_p(t) = \frac{a}{c} \left(\frac{b \sin (c\ (t-{t_0})+d)}{32 E (b \cos (c\ (t-{t_0})+d)-J \hat{\ep})}+\frac{2 J \hat{\ep}\ \tanh ^{-1}\left(\frac{(b+J \hat{\ep})}{4\sqrt{2 E}} \tan \left(\frac{1}{2} (c\ (t-{t_0})+d)\right)\right)}{\left(32E\right)^{3/2}}\right) + f^0_p.
\end{equation}
In Section \ref{sec-bulk}, where we will discuss interpretation of the perturbed SYK in terms of bulk dual, we will argue that this solution corresponds to a horizonless geometry ({hence the subscript $p-$ for ``Poincare''$-$  in the $f(t)$ in Eq., \eq{eq:fpp})}.

$\bullet$ Case II: For $E>0$ (in this case $c$ and $d$ become purely imaginary)
\begin{equation}\label{eq:fbhp}
 f_{bh}'(t) =  \frac{a}{\left(b \cosh\left(|c|\ (t-{t_0}) + |d|\right)-J \hat{\ep}\right)^{2}},
\end{equation}
this gives us
\begin{equation}
 f_{bh}(t) = \frac{a}{|c|} \left(\frac{b \sinh (|c|\ (t-{t_0})+|d|)}{32 E (b \cosh (|c|\ (t-{t_0})+|d|)-J \hat{\ep})}+\frac{2 J \hat{\ep}\ \tan ^{-1}\left(\frac{(b+J \hat{\ep})}{4\sqrt{2E}} \tanh \left(\frac{1}{2} (|c|\ (t-{t_0})+|d|)\right)\right)}{\left(32E\right)^{3/2}}\right) + f^0_{bh}.\label{fbh}
\end{equation}
In Section \ref{sec-bulk}, we will interpret this solution to correspond to a black hole geometry ({hence the subscript $bh$ in the $f(t)$ in Eq., \eq{eq:fbhp})}.

$\bullet$ Case III: $E=0$ (we take $E = \pm J^2 e$ with $e>0$ and then we take the limit $e \to 0$)

The constants \eq{abcd} up to leading order in $e$ are
\begin{align}
 &a = (4 J e)^2,\quad b= \left(J \hat{\ep} \pm \frac{16 J e}{\hat{\ep}}\right), \quad c = J \sqrt{\f{\mp e}{2}},\quad d=\left(2 x_0\sqrt{\f{2}{\hat{\ep}}}\right) \sqrt{\mp e},\quad x_0^2 =\left( e^{-\phi_0/2} - \f{4}{\hat{\ep}}\right).
\end{align}
When we substitute above values in \eq{ephi} and consider   
\begin{equation}\label{E0lim}
 \s{\f{e}{2}} \left({J t} + \f{4 x_0}{\hat{\ep}}\right) \ll 1,
\end{equation}
then we get
\begin{equation}\label{fp0}
 f'_{E=0}(t) = {\hat{\ep}^2}{\left[4 + \left(\f{J t \hat{\ep}}{4} + \sqrt{\hat{\ep}}\ x_0\right)^2\right]^{-2}}.
\end{equation}
Integrating above equation gives us
\begin{equation}\label{f0}
 f(t)_{E=0}(t) = \left(\f{\hat{\ep}}{4J}\right)\left[\f{8(4 x_0 \sqrt{\hat{\ep}}+J t \hat{\ep})}{64 + (4 x_0 + J t \sqrt{\hat{\ep}})^2 \hat{\ep}} + \tan^{-1}\left(\f{J t \hat{\ep}}{8} + \f{\sqrt{\hat{\ep}}x_0}{2}\right)\right] + f^0.
\end{equation}
Armed with the above analysis of the general solution, we can now present the specific solutions in the two real time segments.

\subsection{The first real time segment $t\in (0,T)$}\label{sec-first}

Continuity of $\phi'(t)$ and $\phi(t)$ at $t=0$ gives
\begin{align}\label{01cont}
 \phi'_0(t)\Big|_{t=0^-} &=  \phi'_1(t)\Big|_{t=0^+} = 0, \nn \\
 \phi_0(t)\Big|_{t=0^-} &=  \phi_1(t)\Big|_{t=0^+} = 2 \log\left(\f{\pi}{\be J}\right). 
\end{align}

 Using the above boundary values (as well as $\lambda=-4J, \hat{\epsilon} = \epsilon_1$) in \eq{EC}, we
get
\begin{align}
  E_1 =  \f{\pi J}{2 \be} \left(\epsilon_{0,cr} - \epsilon_1\right),
  \quad \epsilon_{0,cr} \equiv \f{4\pi}{\be J}.
\label{eps-0-cr}
\end{align}
{See that for $\epsilon_1 > \epsilon_{0, \rm{cr}}$ we have $E_1< 0$. Such solutions correspond to the $f(t)$ given in Eq., \eq{fp} and, as we will see, these correspond to the horizonless solutions in the bulk \cite{Kourkoulou:2017zaj}}\footnote{Note that in this time segment $\be$ should not be interpreted as inverse temperature anymore.} \footnote{We have a difference of a factor
  of 4 with \cite{Kourkoulou:2017zaj}.}. In this range of $\epsilon_1$,
therefore, we have the oscillatory solutions
\begin{equation}\label{osci-sol}
 \phi_1(t) = - 2\log \left[\left(\f{J}{4E_1}\right)\left(\s{J^2\ep_1^2+ 32 E_1} \cos(\s{\f{-E_1}{2}}\ t) - J \ep_1\right)\right].
\end{equation}
As $\epsilon_1$
increases, $E_1$ decreases till $\epsilon_1 = (8\pi/\be J) $, when $E_1$ reaches
its minimum possible value ({\it cf.} Eq. \eq{h-min}),
\begin{align}
  V_{min} = - (J^2 \ep_1^2/32) = -(2\pi^2/\be^2).\label{v-min}
\end{align}
At this point, $\phi$ is at
the bottom of the potential, hence it is given by a constant:
\begin{align}
 e^{\phi_1(t)} &=  f_1'(t)=\left(\f{\pi}{\be J}\right)^2,\label{phi1} \\ 
 f_1(t) &=  \left(\f{\pi}{\be J}\right)^2 (t- T). \label{f1p}
\end{align}
Integration constant in $f_1(t)$ is chosen to ensure the continuity of $f(t)$ at $t=T$. 

\subsection{The final real time segment $(t>T)$}\label{sec-final}

The boundary condition for the solution $\phi(t)$ in this segment is given
by the continuity of $\phi'(t)$ and $\phi(t)$ at $t=T$:
\begin{align}
 \phi_2(t)\Big|_{t=T^-} &=  \phi_1(t)\Big|_{t=T^+},  \nn \\
 \phi'_2(t)\Big|_{t=T^-} &=  \phi'_1(t)\Big|_{t=T^+}.
 \label{matching-at-T}
\end{align}
Using \eq{osci-sol} we can write down $E_2$ (with $\hat{\ep} = \ep_2$)
\begin{align}\label{E-2}
 E_2 &= \f{\phi'_1(T)^2}{2} + 2 J^2 e^{\phi_1(T)} -\ep_2\  \f{J^2}{2} e^{\phi_1(T)/2}\nn \\ &= \f{J^2 e^{\phi(T)/2}}{2} \left[\left(\f{\phi'_1(T)^2\ e^{-\phi_1(T)/2}}{J^2} + 4 e^{\phi_1(T)/2}\right) - \ep_2\right]\nn \\
 & = \f{J^2 e^{\phi(T)/2}}{2}\ (\ep_{cr}-\ep_2)~.
\end{align}
Note that in the above, we have defined, 
\begin{equation}
\ep_{cr}=  \f{(\phi'(T))^2 + 4 J^2 e^{\phi(T)}}{ J^2 e^{\phi(T)/2}}.
\label{ep-cr-general}
\end{equation}
We chose $\ep_1$ such that $E_1<0$. Now we are interested in the solutions where $E_2>0$. This will happen for $\ep_2 < \ep_{cr}$. Let us proceed with the simple solution \eq{phi1} for the first line segment. 
\begin{equation}
 e^{\phi(T)} =  \left(\f{\pi}{\be J}\right)^2,\quad  \phi'(T) = 0. 
\label{static}
\end{equation}
In this case 
\begin{align}
  E_2 =  \f{\pi J}{2 \be} \left(\epsilon_{cr} - \epsilon_2\right),
  \quad \epsilon_{cr} \equiv \f{4\pi}{\be J}.
  \label{ep-cr-special}
\end{align}
Using \eq{abcd}, ${a_2,b_2,c_2,d_2}$ can be written in terms of $E_2$ and $\epsilon_2$. The solution in the final time segment, $t>T$ is
\begin{equation}\label{phi2}
e^{\phi(t)} = f_2'(t)= \frac{a_2}{\left(b_2 \cosh\left(|c_2|\ (t-{T})\right)-J \ \ep_2\right)^{2}},  \hspace{6.8cm}
\end{equation}
\begin{equation}
 f_{2}(t) = \frac{a_2\ b_2 \sinh (|c_2|\ (t-{T}))}{|c_2|\left(b_2^2-J^2\ \ep_2^2\right) (b_2 \cosh (|c_2|\ (t-{T}))-J \ \ep_2)}, \hspace{6cm}  \nn \\ \hspace{5cm}+\ \frac{2 a_2\ J \ \ep_2\ \tan ^{-1}\left(\frac{(b_2+J \ \ep_2)}{\sqrt{b_2^2-J^2 \ \ep_2^2}} \tanh \left(\frac{1}{2} (|c_2|\ (t-{T}))\right)\right)}{|c_2|\left(b_2^2-J^2 \ \ep_2^2\right)^{3/2}} \label{f2bh}
\end{equation}
Here $a_2 = (4E_2/J)^2$, $b_2= (J^2\ep^2_2 + 32E_2)^{1/2}$, $c_2 =
(-E_2/2)^{1/2}$, and to ensure continuity of $f(t)$ the integration constant in $f_2(t)$ is set to $0$. In the final real time segment, $t>T$, we chose $0 < \epsilon_2
< \left({4\pi}/{\be J}\right)$. This choice makes $E_2 >0$ and we have
a scattering solution (as can be seen from \autoref{potential}). We
would like to show that this choice makes certain observables in SYK
theory to thermalize at large times, this could be demonstrated by computing large time values of 1-point and
2-point functions. The dual process to thermalization in the field
theory is black hole formation in the bulk gravity theory (see below).

\subsection{Qualitative understanding of the critical value
  $\ep_2=\ep_{cr}$\label{sec-qual-cr}}

It is important to comment upon the  qualitative understanding of the
various behaviours of the final trajectory. It is easy to see with the help of Figure
\ref{fig-ep-cr}. When we change the value of $\hat \ep$ from $\ep_1$
to $\ep_2$, the shape of the potential curve charges. The matching
condition \eq{matching-at-T} implies the following graphical
construction. For simplicity, let us assume that $\phi'_1(T)=0$; then
$\phi_2'(T)=0$. Combining with the condition that $\phi_1(T)=
\phi_2(T)$, the new trajectory after the quench is given by the
following simple construction: take the vertical line $\phi= \phi(T)$;
the point of intersection of this line with the new potential curve
(for $\ep_2$) defines the turning point (since $\phi_2'(T)=0$) of the
new trajectory.

\begin{figure}[H]
\begin{center}
  \includegraphics[scale=1]{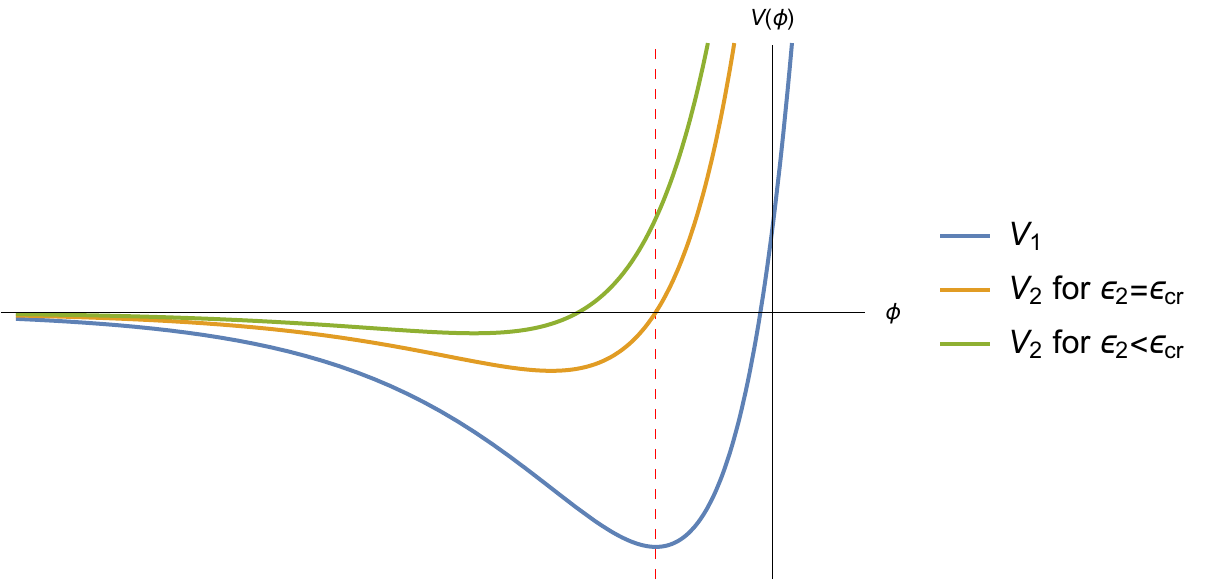}
  \caption{We have taken here $\ep_1= \f{8\pi}{\beta J}$ for
    simplicity. With this value, the $\phi$-particle sits at the
    bottom for all times $0<t<T$, with the value of $\phi$ given by
    $\phi_1$ in \eq{phi1}. At $t=T$, we perform the quench $\ep_1 \to
    \ep_2$, which modifies the shape of the potential. Depending on
    the value of $\ep_2$, the line $\phi=\phi_1$ may hit the potential
    curve at the right turning point below the $\phi$-axis (in which
    case the geometry remains horizonless) or above (in which case the
    geometry develops a horizon). At the critical value $\ep_2
    =\ep_{cr}$, \eq{ep-cr-special}, the shape of the potential is such
    that the line $\phi=\phi_1$ hits the potential curve exactly on
    the $\phi$-axis. For $\ep_2 < \ep_{cr}$, this intersection point
    rises above the $\phi$-axis, corresponding to the formation of a
    black hole. }
\label{fig-ep-cr}
\end{center}
\end{figure}

In Figure \ref{fig-ep-cr}, we have explained how this construction
exhibits the critical value $\ep_2=\ep_{cr}$. In Section \ref{sec-choptuik}
we will explain how this critical value, which demarcates
the boundary between formation of black holes and otherwise, corresponds
to a Choptuik-like phenomenon. 

\section{Interpretation in terms of bulk geometry}\label{sec-bulk}
{It is now well known that the SYK model, being solvable at strong coupling, is a simple model of holography. The low energy bulk dual to this model can be presented either as Polyakov-gravity \cite{Mandal:2017thl} } 
\begin{equation}
 S_{pol} = \f{1}{16\pi b^2}\int_{\Gamma} d^2x\s{g}\ \left[R\f{1}{\Box}R -16 \pi \mu\right] + \f{1}{4\pi b^2}\int_{\del \Gamma} dt\s{\gamma}\ \mathcal{K}\f{1}{\Box}R + \f{1}{4\pi b^2} \int_{\del \Gamma} \s{\gamma}\ \mathcal{K}\f{1}{\Box} \mathcal{K}
\end{equation}
or Jackiw-Teitelboim (JT) gravity \cite{Maldacena:2016upp} 
\begin{equation}
  S_{JT} =- \f{\Phi_0}{16 \pi G}
  \left(\int_M\ d^2x\ \s{g} R  + 2 \int_{\partial M} dt \s{\gamma} \mathcal{K} \right) -  \f{1}{16 \pi G}\left(\int_M\ d^2x\ \s{g}\Phi (R + 2) + 2 \int_{\partial M} dt \s{\gamma} \Phi (\mathcal{K}-1) \right).
  \label{JT-main}
\end{equation}
The deformation of the boundary Hamiltonian with $H^{(s)}_M$ corresponds to adding some matter in the bulk along with the pure gravity. We assume that this matter only couples to the metric and not to the Liouville field/dilaton($\Phi$). For simplicity we will restrict the discussion to JT gravity in this work, although  it is straightforward to extend our analysis to the Polayakov gravity. In the case of JT gravity the total action is 
\begin{equation}
  S = S_{JT} + S_{\text{matter}} + S_{bdry}.
  \label{JT}
\end{equation}

We discuss the bulk equation of motion and matter stress tensor in the subsequent sections. The pseudo Nambu-Goldstone mode, $f(t)$ is represented in the bulk dual as large diffeomorphism of asymptotically AdS$_2$ metric \cite{Mandal:2017thl} or equivalently as the shape of the boundary curve \cite{Maldacena:2016upp}. The equivalence of the two view points is explained in section \ref{sec-collapse}.

\subsection{Horizon dynamics in terms of the saddle point value of $f(t)$}

As emphasized in \cite{Kourkoulou:2017zaj}, the relation between saddle point solutions of $f(t)$ and bulk geometry is established by the fact that the shape of the UV boundary curve $z \sim \delta f'(t)\sim \delta e^{\phi(t)}$ is determined by the solution for $f$ or $\phi$. In particular, for a horizon to exist, there must be points where $f'(t)$ vanishes so that $z$ reaches zero. In case of the unbounded
motion such as \eq{fbh} (with $E>0$), $\phi(t)$ eventually reaches
$-\infty$, hence the geometry has a horizon. On the other hand, for
the bounded motion \eq{fp} (with $E<0$), the geometry is horizonless
(it has an oscillating boundary curve, given by the specific form of
$f_p'(t)\not= 0$).

In the first quench, the shape of the potential curve changes from Fig \ref{potential} (b) $\to$ (a). The energy perturbation in this case is negative (as is easy to see from the figure), i.e. energy is drawn out from the system. The nature of motion changes from unbounded to bounded; hence, as explained above, horizon disappears. The apparent violation of second law of black hole thermodynamics is permitted because of the negative energy perturbation (which, in the bulk, represents negative energy of matter). In the second quench, the shape of the potential changes from blue curve in Fig \ref{fig-ep-cr} to the green curve; horizon is now created since the motion goes from bounded to unbounded.  From the Fig \ref{fig-ep-cr}, it is clear that in going from blue curve to the green one, positive energy is pumped into the system. This corresponds to normal matter and the formation of the horizon can be identified with gravitational collapse into a black hole.

\subsubsection{Topology change\label{sec-topology}} It is important to note that formation of a
horizon is related to a qualitative change in the nature of the
saddle point solution for the map $f$. For horizonless solutions such
as AdS$_2$-Poincare, $f$ maps $R\to R$: $f(t) \propto t$. However for
horizons to appear, $f'(t)$ must vanish; which implies vanishing of
the Jacobian of the map $t \to f(t)$. This, in fact, changes the nature of the
saddle point from $f:R\to R$ to $f: R \to I$ with, e.g. $f(t) \propto
\tanh(\pi t/\beta)$, where $I$ denotes the interval $[-1, 1]$.

\subsection{Choptuik phenomenon \footnote{We
    mean a Choptuik-like phenomenon. See
    footnote \ref{ftnt-choptuik}.}}\label{sec-choptuik}

We found above, that for quenches satisfying $\epsilon_2 <
\epsilon_{cr}$, or alternatively $\Delta \ep > \Delta \ep_{cr}$ lead
to the formation of a black hole geometry (where $\Delta\ep = \ep_1 -
\ep_2, \Delta\ep_{cr}= \ep_{cr}- \ep_1$) (see Section
\ref{sec-qual-cr}, in particular Figure \ref{fig-ep-cr} for a
qualitative understanding). We will soon discuss the interpretation of
this phenomenon in terms of gravitational collapse.  Below we would
like to characterize the black hole solution, thus formed, by
identifying its temperature.  From \eq{phi2}, we see that at long times
the $\cosh\left(|c_2|(t-T)\right)$ term dominates in the
denominator. Thus at large times $f'$ is of the asymptotic form
\begin{align}
  f'(t) \sim A \sech^2\left(|c_2|(t-T)\right).
  \label{fp-late}
\end{align}
We identify the first term with a thermal result at inverse
temperature $\b_{bh}$ by demanding periodicity under
an imaginary time shift by $\beta_{bh}$, i.e. under $t
\to t+ i\beta_{bh}$. This gives $|c_2|\beta_{bh}= \pi$.
By using $|c_2|= \sqrt{E_2/2}$ and Eqs \eq{E-2} and \eq{static}, 
we can get 
\begin{align}
  \f\pi{\be_{bh}} =  \f12\sqrt{\f{\pi J}{\beta}}\ \sqrt{\ep_{cr}- \ep_2}.
  \label{beta-bh}
\end{align}
Here $\beta \equiv 2l$ is related to the initial $|B_s(l)\ran$.  A
useful check of the above formula is to note that for $\ep_2=0$,
$\beta_{bh}= \beta$, as expected (using $\ep_{cr} =4\pi/\be
J$).\footnote{Essentially, with the choice of the static solution
  \eq{static} during the first phase of evolution from $t=0$ to $t=T$,
  the state throughout remains $|B_s(l)\ran$, including immediately
  after the second quench. If one performs a sudden quench to
  $\ep_2=0$, the state remains $|B_s(l)\ran$, while the Hamiltonian
  comes back to $H_{SYK}$, for which $|B_s(l)\ran$ of course has an
  effective temperature $\be = 2l$.\label{ftnt-betabh-beta}}

The above formula \eq{beta-bh} shows that the temperature of the black
hole formed by gravitational collapse has a non-analyticity
in $\epsilon$:
\begin{align}
  T_{bh} \propto  \sqrt{\ep_{cr}- \ep_2} 
\label{t-bh}
\end{align}
What we have here is akin to the Choptuik phenomenon, discovered in
\cite{PhysRevLett.70.9}, where gravitational collapse of spherical
matter was considered. The initial data was parameterized by a
parameter $p$. It was found that there exists a critical value
$p_{cr}$ such that black holes form only when $p>
p_{cr}$. Furthermore, the mass of the resulting black hole is given by
\begin{align}
M_{bh} \sim (p - p_{cr})^\gamma.
\label{M-bh-crit}
\end{align}
Here the quench to $\ep_2$ represents the one-parameter space of
perturbations to the system at $\ep_1$. In case $\ep_2 > \ep_{cr}$ the
change $\Delta\ep = \ep_1 - \ep_2$ is less than $\Delta\ep_{cr} \equiv
\ep_1- \ep_{cr}$ and the perturbation is too weak to form a black
hole. For $\Delta\ep > \Delta\ep_{cr}$ the perturbation is strong enough
to form a black hole; the relation \eq{t-bh} can be written
as
\[
T_{bh}\propto (\Delta\ep - \Delta\ep_{cr})^{\f12}.
\]
We note that in case the solution in the first phase has negative
energy $E_1<0$ but $\phi$ is not at the bottom of the potential, the formula for
the critical value of $\ep_2$ is more general
\[\ep_{cr}=  \f{(\phi'(T))^2 + 4 J^2 e^{\phi(T)}}{ J^2 e^{\phi(T)/2}}.
\]
A couple of comments are in order.

\begin{itemize}

\item
If we keep track of subleading terms of \eq{phi2} in the large time
expansion, we get 
\begin{align}
  f'(t) \xrightarrow{t-T \to \infty} A e^{-i \sqrt{E_2} (t-T)}
  +  B e^{-i \f32 \sqrt{E_2} (t-T)} + \hbox{higher transients}.
\label{later-fdot}
\end{align}
We note that while the leading term does have the periodicity $t\to
t+i\beta_{bh}$, the subleading term is periodic under $t \to t+
2i\beta_{bh}$. We read off thermal behaviour and the effective
temperature from the leading term.

\item
  We can get the behaviour \eq{t-bh} more qualitatively, as follows.
  The advantage of this method is that it works for arbitrary $q$ (not
  necessarily $q=4$). To work this out, note that for very late times,
  $ t-T \to \infty$, $\phi(t) \to -\infty$ hence
  $V(\phi(t)) \to 0$. In particular, at sufficiently large $t$, we
  have $E_2 \gg V$. Eq \eq{quad} now becomes
\begin{equation*}
  -t = \int_{\phi_0}^{\phi} d\phi'\left(\frac{1}{\sqrt{2 (E_2 - V(\phi))}}\right)
  \approx \int_{\phi_0}^{\phi} d\phi'\left(\frac{1}{\sqrt{2 (E_2)}}\right).
\end{equation*}
Hence
\begin{equation*}
 \phi \approx \phi_0-\sqrt{2 E_2}\ t, 
\end{equation*}
and 
\begin{equation}
  f'(t) = e^{\phi}  \Rightarrow e^{-\sqrt{2 E_2}\ t}. 
\label{fdot-asym-a}
\end{equation}
This is of the same form as the leading term of \eq{later-fdot};
hence we get the same critical behaviour for the black hole temperature.

\item
Note that in the other phase, $\ep_2 > \ep_{cr}$ the solution for
$f(t)$ is oscillatory and $f'(t)$ never goes to zero; hence the
solution does not have a horizon.\footnote{Note that in this domain
  too, there a criticality in the time period $\mathcal{T}$ of the bounded
  motion, which diverges in the limit $\ep_2 \to \ep_{cr}^{+}$ as 
$ \mathcal{T} \sim (\ep_2-\ep_{cr})^{-\f12}$.
The physical significance of this criticality is not clear.
\label{ftnt-time-of-flight}}

\end{itemize}

\subsubsection{Reason for the Choptuik phenomenon\label{sec-reason}}
It is important to note that the Choptuik phenomenon arises because of
the mass gap mentioned in Section \eq{sec-mass-gap}. E.g.  if one
starts in the first phase $t\in (0,T)$ in the classical vacuum of the
potential for $\ep_1$, then there is a minimum amount of energy that
must be supplied to the system to reach $E>0$, i.e.  to form a
black hole. In terms of the bulk dual, in the entire range $\ep_2>
\ep_{cr}$, the energy supplied through the change $\ep_1 \to \ep_2$ is
not enough. The geometry of the vacuum configuration remains an
AdS$_2$-Poincare geometry (with the special boundary insertion at the
corner of the Poincare wedge) throughout this range (although the
matter and dilaton changes, causing the lowering of
energy)\footnote{We thank D. Anninos for a discussion on this
  point.}. It is only when $\ep_2 < \ep_{cr}$, or in terms of the
change, $\Delta\ep > \Delta\ep_{cr}$, that the geometry starts
changing-- a horizon develops and grows with increasing $\Delta\ep$.

\subsection{Bulk description of gravitational collapse\label{sec-collapse}}

In this subsection we will derive  near boundary expression for the bulk matter stress tensor, which corresponds to the saddle point value of $f(t)$.

The equations of motion from  action \eq{JT} are
\begin{subequations}\label{jteom}
 \begin{eqnarray}
 R+2 = 0, \hspace{4.1cm}  \\
 \nabla_\mu \nabla_\nu \Phi - g_{\mu\nu} \nabla^2\Phi + g_{\mu\nu}\Phi + T^{\text{M}}_{\mu\nu} =0,
\end{eqnarray}
\end{subequations}
where $T^{\text{M}}_{\mu\nu}$ is the stress tensor corresponding to the bulk matter. 

\subsubsection*{Two equivalent bulk representations of the soft mode}
In this section we will outline two equivalent bulk points of view to represent the pseudo Nambu-Goldstone mode $f(t) $ described before.

In the \textbf{\textit{first}}, we choose a coordinate system such that the boundary curve is always given by $(t,z) = (u,\delta)$ and the metric has an asymptotically AdS$_2$ form \cite{Mandal:2017thl}.
\begin{equation}\label{aads2}
 ds^2 = \f{dz^2}{z^2} + \f{dt^2}{z^2}\left(1+\f{z^2}{2}\{f(t),t\}\right)^2.
\end{equation}
This metric depends on an arbitrary function $f(t)$ and it satisfies $R = -2$. The boundary condition on the dilaton is  $\Phi(u,\delta) = \Phi_r(u)/\delta  + O(\delta)$, where $\Phi_r(u)$ is a given boundary condition. Metric \eq{aads2} is a large diffeomorphism of the Poincare AdS$_2$ metric \cite{Mandal:2017thl}.

In the \textbf{\textit{second}}  point of view, the coordinate system is chosen such that the metric is Poincare $AdS_2$ and we define the boundary curve as $(t,z) = (t(u), z(u))$ (the function  $u \mapsto t(u)$ used here is the same as the function $t \mapsto f(t)$ used in the above paragraph). Further, we 
impose two conditions,
(a) the induced metric on the boundary is $\gamma_{uu}=1/\delta^2$, which gives $z(u)= \delta\ t'(u) + O(\delta^3)$ and (b) $\Phi (t(u), \delta\  t'(u))= \Phi_r(u)/\delta  + O(\delta)$ where $\Phi_r(u)$ is a given boundary condition for the dilaton. 
In this point of view, the physical placement of the boundary curve  is parameterized by $t(u)$.

\subsubsection{Profile of the dilaton and matter stress tensor\label{sec-shell}}
In this section we will take the second point of view, then (\ref{jteom}b) is 
\begin{subequations}\label{Tmn}
\begin{align}
 \del_z\left[ \del_z\Phi(t,z) + \f{\Phi(t,z)}{z}\right]+ T_{tt}^M(t,z) =0,\label{Ttt}\\
 \del_t\left[ \del_z\Phi(t,z) + \f{\Phi(t,z)}{z}\right]+ T_{tz}^M(t,z) =0, \label{Ttz}\\
 \del_t^2\Phi(t,z) + \f{1}{z}\left[ \del_z\Phi(t,z) + \f{\Phi(t,z)}{z}\right] + T_{zz}^M(t,z) =0. \label{Tzz}
\end{align}
\end{subequations}
When $T^M_{\mu\nu}=0$, we know the homogeneous solution for \eq{Tmn}, which is 
\begin{equation}
 \Phi_0(t,z) = \f{a + b\ t + c\ (t^2 + z^2)}{z}
\end{equation}
where $a,b$ and $c$ are arbitrary constants. From \eq{Ttz}, we have 
\begin{equation}
 \del_z \Phi(t,z) + \f{\Phi(t,z)}{z} + \f{1}{\del_t} T^M_{tz}(t,z)=0.
\end{equation}
Combining this with \eq{Tzz}, we get
\begin{equation}\label{Dil}
 \Phi(t,z) = \f{1}{z} \left[\left(\f{1}{\del^3_t}\right) T^M_{tz}(t,z) - z \left(\f{1}{\del^2_t}\right) T^M_{zz}(t,z)\right] + \Phi_0(t,z).
\end{equation}
The equations above should be understood in the momentum space. \par In order to keep proper length of the boundary fixed, we impose
\begin{equation}
 g_{uu} = \f{1}{\delta^2}, \quad g_{uu} = \f{t'(u)^2 + z'(u)^2}{z(u)^2} \implies z(u) = \delta\ t'(u) + O(\delta^3).
\end{equation}
The boundary condition on the dilaton is
\begin{equation}\label{Dil-BC}
\Phi( t(u), z(u)) = \Phi( t(u),\delta\  t'(u)) = \f{\bar{\Phi}_r}{\delta} + O(\delta) ,
 \end{equation} 
here $\bar{\Phi}_r$ is a constant (We have chosen boundary condition for dilaton to be independent of the boundary time). Let us assume a series expansion of $T^M_{tz}$ and $T^M_{zz}$
\begin{equation}\label{T-exp}
 T^M_{tz} = g_0(t) + g_1(t)\ z + \cdots \nn \\
 T^M_{zz} = h_0(t) + h_1(t)\ z + \cdots
 \label{Texp}
\end{equation}
Plugging this in \eq{Dil}, we get
\begin{align}\label{Dil-exp}
 \Phi(t,z) &= \f{1}{z}\left[\left(\f{1}{\del^3_t}\right) g_0(t) + a + b\ t + c\ t^2\right] + c\ z + \sum_{n=0} z^n \left[\left(\f{1}{\del^3_t}\right) g_{n+1}(t) - \left(\f{1}{\del^2_t}\right) h_n(t)\right].
\end{align}
Imposing \eq{Dil-BC} on \eq{Dil-exp}, we get
\begin{equation}
 \tilde g_0(t(u)) = \f{\bar{\Phi}_r\ t'(u)}{ a + b\ t(u) + c\ t(u)^2 } ,\quad \text{here}\ \left(\f{1}{\del^3_t}\right) g_0(t) = \tilde g_0(t)
 \label{gt-g}
\end{equation}
and 
\begin{subequations}
\begin{align}
 t(u) &= t_1(u)\ \Theta(u-T) + t_2(u)\ \Theta(T-u) + t_0, \label{t-profile}\\
 t'(u) &= t'_1(u)\ \Theta(u-T) + t'_2(u)\ \Theta(T-u) + (t_1(u)-t_2(u))\delta(u-T). \label{t'-profile}
\end{align}
\end{subequations}
where $t_0 = T\left({\pi}/{\be J}\right)^2$ (this choice makes $t(0) = 0$)now we rely on $nAdS_2/nCFT_1$ duality to identify the reparameterization mode $t(u)$ with the field theory solutions,$f(t)$ (we have shifted the solution by a constant which is still a symmetry of the system), for the second and final time segments, which were obtained in \autoref{app-onesidsol}. For notational consistency take $t(u) \equiv f(t)$
\begin{align}
 t_1(u) &= \left(\f{\pi}{\be J}\right)^2 (u- T),\nn\\
 t_2(u) &= \f{\pi}{\be J^2} \tanh\left(\f{\pi}{\be}(u-T)\right)  + \f{\ep_2}{8 J} \left[ 2 \tan^{-1}\left(\tanh\left(\f{\pi}{\be}(u-T)\right)\right) -\tanh\left(\f{\pi}{\be}(u-T)\right)\right] \nn \\ &\hspace{1.5cm}+ \f{\ep_2}{8 J}\ \sech^2\left(\f{\pi}{\be}(u-T)\right)\left[\left(\f{\pi}{\be}(u-T)\right) + \sinh\left(\f{\pi}{\be}(u-T)\right)\right].
\end{align}
$t_2(u)$ is the small $\ep_2$ expansion of \eq{f2bh}. From above equations we can see that $\tilde g_0(t(u))$ has discontinuity/ singularity at $u=T$, which will also reflect in expression for \eq{Dil} and \eq{Tmn} at the boundary. Extension of this discontinuity/singularity into the bulk gives the picture of an infalling shell of matter.\\
An important aspect of the above derivation is that one could glean
some aspects of the matter stress tensor from pure consistency
requirements in a model-independent way, i.e. without mentioning
specifically what is the matter field dual to the perturbation $H_M^{(s')}$.
\vspace{-.5cm}
\subsubsection*{$\ep_2 = 0$ case}
\vspace{-.5cm}
\begin{figure}
        \centering
        \begin{tikzpicture}[scale = 1, every node/.style={scale=1}]
        \draw [fill=green] (1,0) -- (4.5,0) -- (4.5,3.5) -- (1,0);
        \draw [dashed,red,ultra thick](0,0)--(4.5,0);
        \draw [fill=pink](0,0) arc [radius=2.25, start angle=180, end angle= 360];
        \draw [red,ultra thick](0,0)--(4.5,4.5);
        \draw  (4.5,0)--(4.5,4.5);
        \draw [fill=red](0,0) circle [radius=.1];
        \draw[thick] (3.04,0) to [out=90,in=268] (3.05,.5) ;
        \draw [dashed,thick] (3.05,.5) to [out=88,in=225] (4.5,4.5);
        \draw [thick] (3.05,.5) to [out=88,in=235] (4.5,3.5);
        \draw [blue](0,0)--(4.5,1.4);
        \node at (2.25,-1) {Euclidean};
        
        \node at (0,3) {$\bullet$ $t(T)$ bulk slice};
        \draw  [->] (.5,2.8) to [out=-45,in = 120](1,.3);
        
        \node at (0,3.5) {$\bullet$ $t(u)$ for $u<T$};
        \draw  [->] (1.2,3.5) to [out=-25,in = 150](3,.3);
        
        \node at (0,4) {$\bullet$ $t(u)$ for $u>T$};
        \draw  [->] (1.2,4) to [out=-25,in = 150](3.5,2.2);
        
        \draw [fill=blue](3.1,.95) circle [radius=.1];
        
        \end{tikzpicture}
        \caption{A cartoon describing the change in the boundary curve. A quantum quench is performed at a boundary time indicated by the blue dot the result of this is that the boundary curve (solid) reaches the $z=0$ surface at a finite Poincare time. This creates a horizon because only the green bulk region can communicate with a boundary observer. In absence of the quench the boundary would have been along the dashed curve and any point in the bulk could communicate with the boundary observer indicating a horizonless geometry.}
        \label{fig:my_label}
    \end{figure}
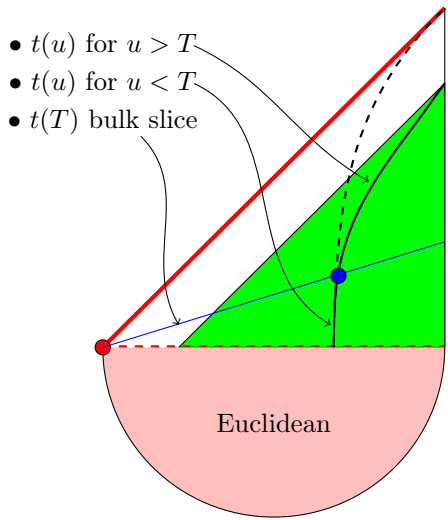
\begin{align}
 t_1(u) &= \left(\f{\pi}{\be J}\right)^2 (u-T), \quad t'_1(u) = \left(\f{\pi}{\be J}\right)^2,\nn \\
  t_2(u) &= \left(\f{\pi}{\be J^2}\right) \tanh\left(\f{\pi}{\be}(u-T)\right),
  \quad t'_2(u)=\left(\f{\pi}{\be J}\right)^2 \sech^2\left(\f{\pi}{\be}(u-T)\right) = \left[\left(\f{\pi}{\be J}\right)^2 - J^2\ t^2_2(u) \right]
\end{align}
Note that $t(T) = t_0$ and $t_1(T)=t_2(T)=0$ hence last term in \eq{t'-profile} vanishes. Now we will evaluate $\tilde g_0(t)$ piecewise
\begin{equation}
 \tilde g_0(t) = 
 \begin{dcases}
  \f{\bar{\Phi}_r}{(\be J)^2} \left(\f{\pi^2}{a + b t + ct^2}\right),\quad t < t_0\ (u<T)\\
  \f{\bar{\Phi}_r}{(\be J)^2} \left(\f{\pi^2 - J^4 \be^2 (t-t_0)^2}{a + b t + ct^2}\right), \quad t>t_0\ (u>T)
 \end{dcases}
 \label{gt}
\end{equation}
and above expression, using \eq{gt-g}, gives us
\begin{equation}
g_0(t) = 
 \begin{dcases}
  -\f{\bar{\Phi}_r}{(\be J)^2}\left(\f{(6 \pi^2 (b + 2 c t_0) (b^2 - 2 a c + 
    2 c t_0 (b + c t_0))}{(a + 
   t_0 (b + c t_0))^4)}\right),\quad t < t_0 \\
  \f{\bar{\Phi}_r}{(\be J)^2} \left(\frac{6 (b+2 c t_0) \left(\beta ^2 J^4 (a+{t_0} (b+c {t_0}))^2-\pi ^2 \left(2 c \left(c {t_0}^2-a\right)+b^2+2 b c {t_0}\right)\right)}{(a+{t_0} (b+c {t_0}))^4}\right), \quad t>t_0.
 \end{dcases}
 \label{g}
\end{equation}
Note that although \eq{gt} is continuous at $t=t_0$, \eq{g} is not. This discontinuity is directly reflected in the expression of matter stress tensor \eq{Texp}, indicating that the matter stress tensor has a discontinuity at $t=t_0$ $(u=T)$, near the boundary.

\section{Correlation functions}\label{sec-2pt}
In this section we will discuss time-evolution of correlation functions
through the process of quantum quench.
\subsection{Non-analyticity in two point function}
We have noted above that when, at some $t=T$, $\hat\ep$ undergoes a
discontinuous (finite) jump, so does $\phi''(t)$ at $T$. This
implies that $\phi'(t)$, or alternatively $f''(t)$ has a kink singularity at $t=T$, see \autoref{fpp}. All lower derivatives of $f$ or $\phi$ are continuous at $t=T$, see \autoref{fig-fp}. We now show that  this singularity reflects a physical
singularity of a 2-point function which, in the presence of the cutoff surface is
\begin{equation}
 G(t_2,t_1) = \frac{({f'}(t_2) {f'}(t_1))^{\Delta}}{|f(t_2)-f(t_1)|^{2\Delta}}
\end{equation}
Let us keep $t_2$ fixed and vary $t_1$ across $t_1=T$. It is easy
to see that the kink singularity of $f''(t)$ corresponds to
a kink singularity of $\del_{t_1} G(t_1, t_2)$.
The singularity in the \autoref{fpp} captures this.

 \begin{figure}[H]
 \centering
\begin{subfigure}[b]{0.4\textwidth}
\centering
 \includegraphics[width=\linewidth]{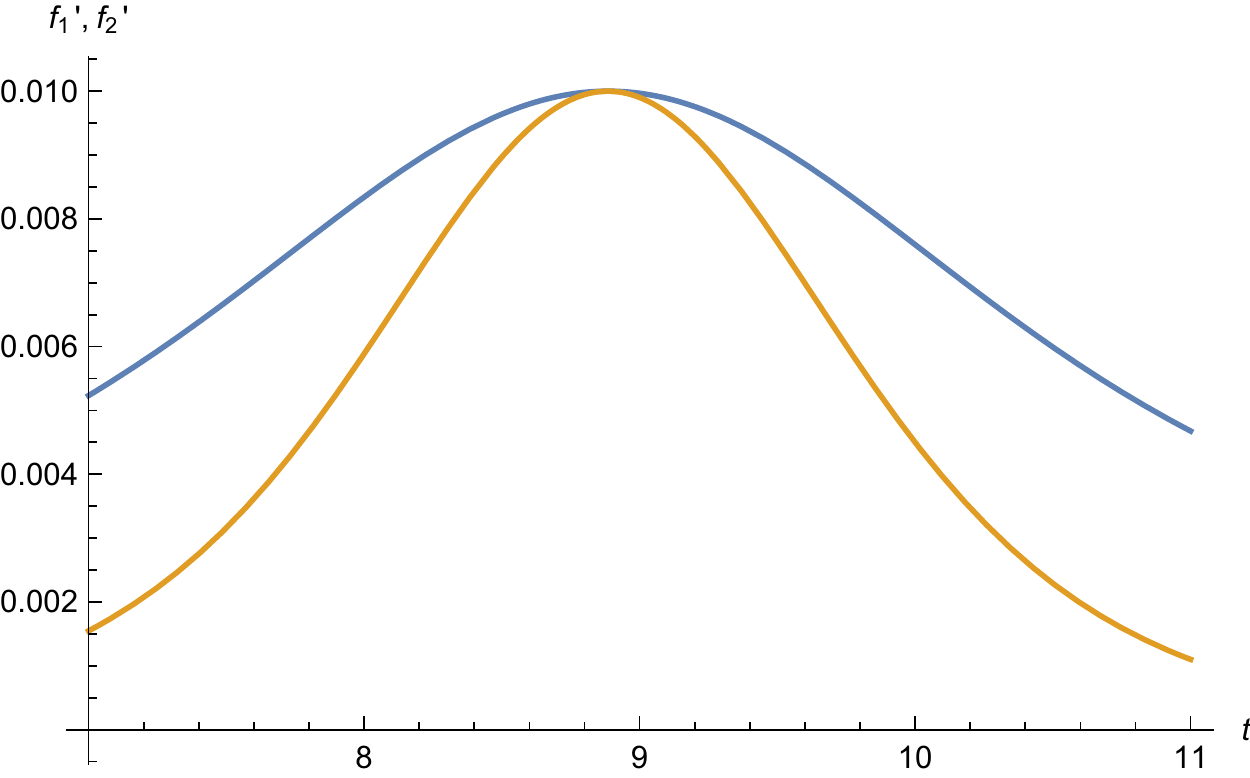}
 \caption{Plot of $f'_1(t)$ (blue) and $f'_2(t)$ (orange)}
\end{subfigure}
\hspace{1cm}
\begin{subfigure}[b]{0.4\textwidth}
\centering
 \includegraphics[width=\linewidth]{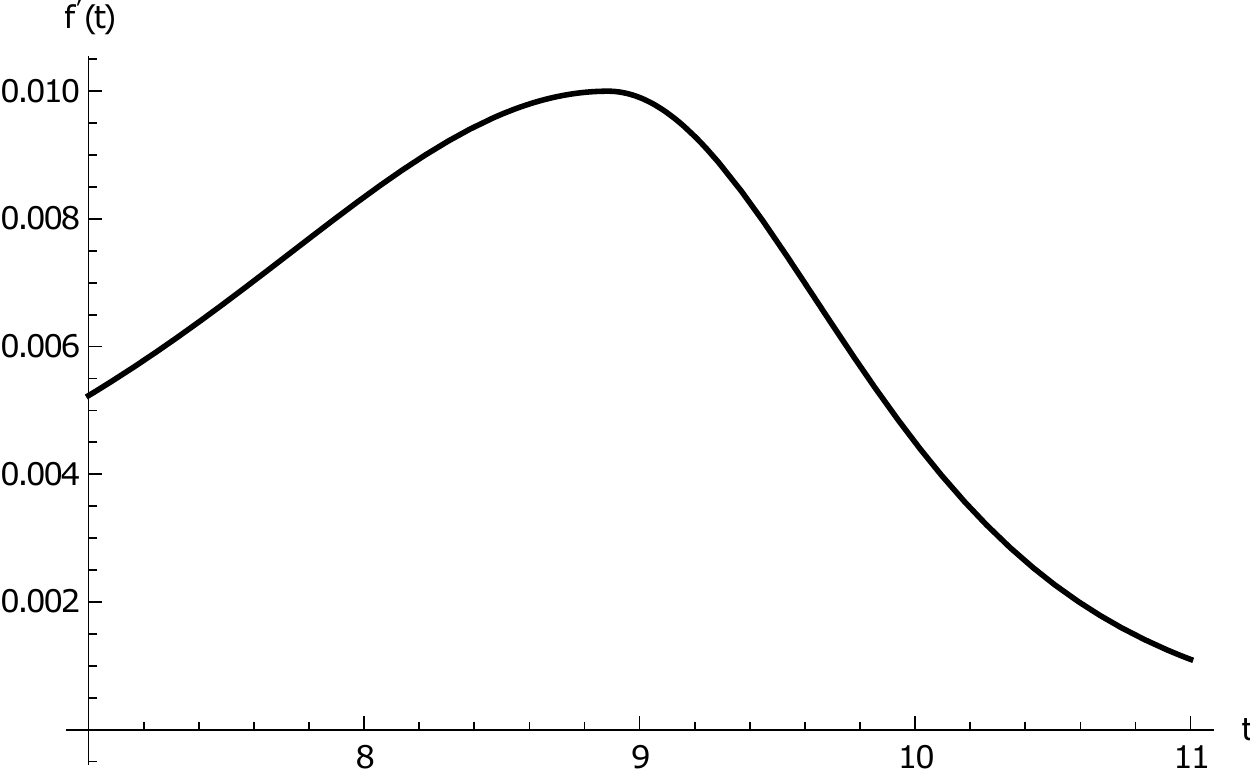}
\caption{Plot of $f'(t)$}
\end{subfigure}
\caption{The derivative $f'(t)$ is given by $f_1'(t)$ for $t<T$ and
  $f_2'(t)$ for $t>T$. The plots of $f'_1(t)$ and $f'_2(t)$ are given in the left panel; these are generated using \eq{osci-sol} and
  \eq{phi2} with $\beta = \pi$, $\e_1=.6$, $\e_2=.2$ and $J=10$.
  $T$ is chosen to be period of $f'_1(t)$; numerically $T=8.8$. The plot
  of $f'(t)$ is taken by combining the relevant segments of $f_1'(t)$ and
  $f_2'(t)$, and is shown in the right panel. Note that the curve is smooth. }
\label{fig-fp}
\end{figure}
 \begin{figure}[H]
 \centering
\begin{subfigure}[b]{0.4\textwidth}
\centering
 \includegraphics[width=\linewidth]{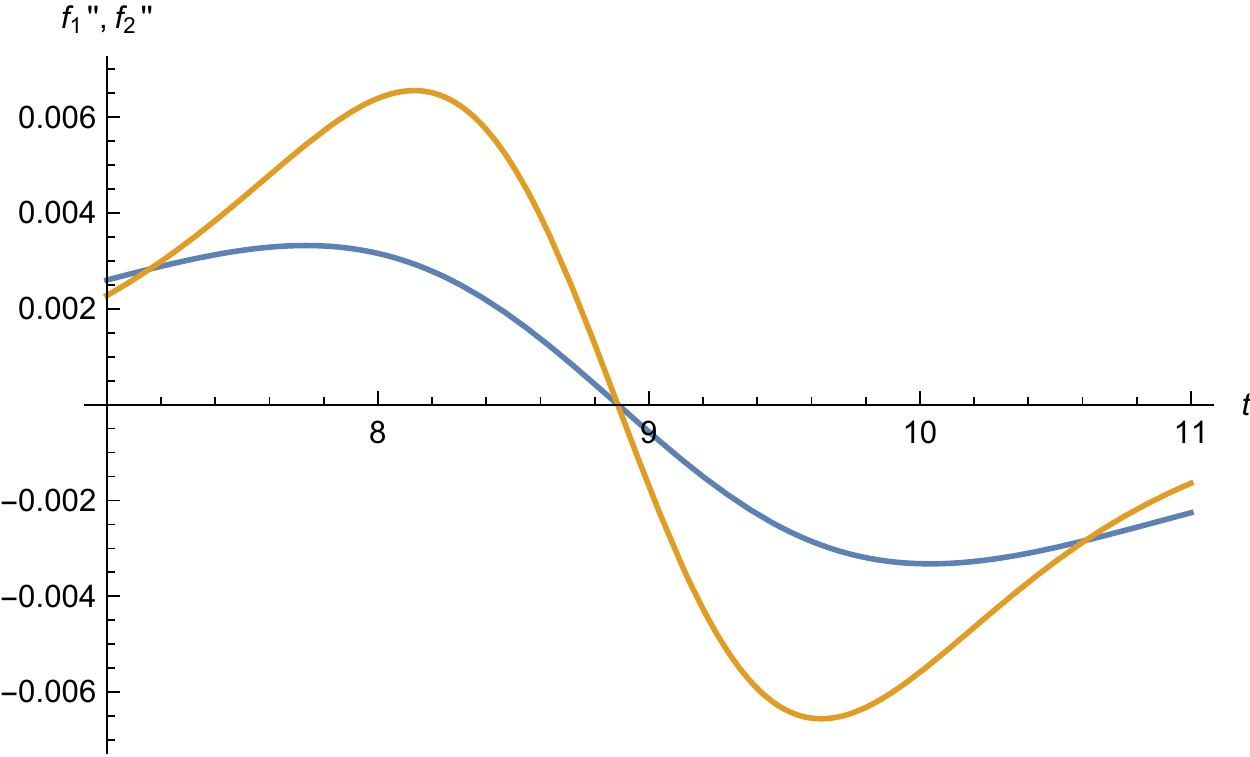}
 \caption{Plot of $f''_1(t)$ (blue) and $f''_2(t)$ (orange)}
\end{subfigure}
\hspace{1cm}
\begin{subfigure}[b]{0.4\textwidth}
\centering
 \includegraphics[width=\linewidth]{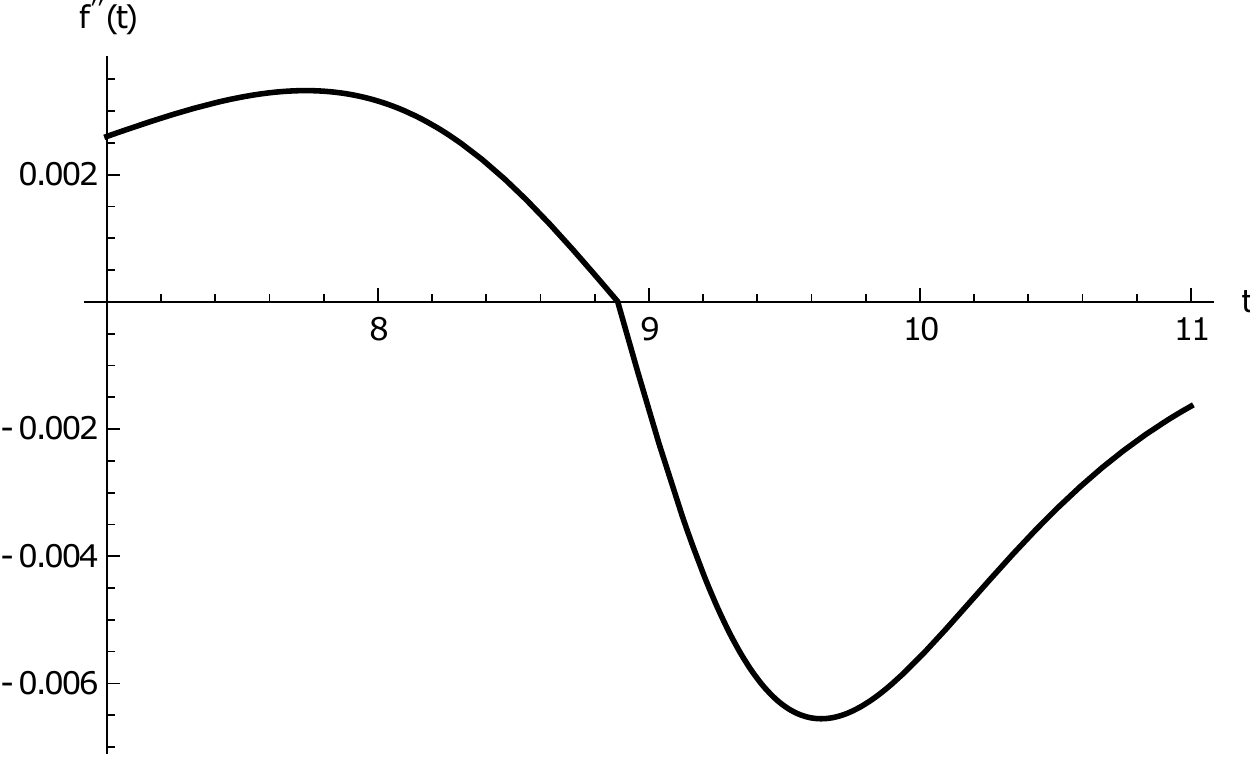}
\caption{Plot of $f''(t)$}
\end{subfigure}
\caption{The derivative $f''(t)$ is given by $f''_1(t)$ for $t<T$ and $f''_2(t)$ for $t>T$. The plots of $f''_1(t)$ and $f''_2(t)$ are given in the left panel; these are generated using the same data as \autoref{fig-fp}. The plot of $f''(t)$ is taken by combining the relevant segments of $f_1''(t)$ and $f_2''(t)$, and is shown in the right panel. Note that the curve has a kink singularity .}
\label{fpp}
\end{figure}

\subsection{Thermalization}\label{sec-thermalization}

In this section we will show that a string of local operators
thermalize in the above described quench protocol \eq{ehat}. We will extract the value of the temperature to which these
operators thermalize.  \par We start by definition of  thermalization for
a string of local operators in an excited state $\ket{\psi}$, it is as follows
\begin{equation}
 \braket{\psi|\mathcal{O}(t + t_1)\cdots\mathcal{O}(t + t_n)|\psi}\quad \xrightarrow{\lim{t\to \infty}} \quad \text{Tr}(\rho_\beta \  \mathcal{O}(t_1)\cdots\mathcal{O}(t_n) ) +\kappa(\beta) \ e^{-\gamma(\beta) t}
\label{thermalization}
\end{equation}
where $\gamma(\beta)$, $\kappa(\beta)$ are temperature dependent constants and $\rho_\beta$ is the equilibrium density matrix. We can evaluate the 1-point and the 2-point functions at large $N$ by simply coupling the reparameterization mode to them and evaluating these on the saddle point solutions \eq{f1p} and \eq{f2bh}.

\subsubsection{Two point function}
The zero temperature 2 point function in the IR limit is
\begin{equation}
  \braket{\psi_i(t_1)\psi_i(t_2)} = G(t_1,t_2) = \frac{C_\Delta}{(t_1-t_2)^{2 \Delta}}, \quad \text{with}\quad C_\Delta = \left[\left(\f{1}{2}-\Delta\right) \f{\tan(\pi \Delta)}{J^2 \pi }\right]^\Delta,\ \Delta = \f{1}{q}. 
\end{equation}
The conformal transformation of the above solution is given by 
\begin{equation}\label{contran}
 G^f(t_1,t_2) = {[f'(t_1) f'(t_2)]^\Delta}\ G(f(t_1),f(t_2)).
\end{equation}
The thermal 2 point function at inverse temperature $\beta$ of the SYK theory ($\hat{\e}=0$) is given by $f(t) = \alpha\ \tanh(\pi t/\beta + c_1) + c_2$ with arbitrary constants $\alpha$, $c_1$ and $c_2 $. This gives us
\begin{equation}
 G_\beta(t_1,t_2) =  \f{C_\Delta}{\left[\f{\beta}{\pi}\ \sinh(\f{\pi (t_1-t_2)}{\beta})\right]^{2\Delta}}.
\end{equation}
We use \eq{f2bh} and \eq{contran} to evaluate the two point function
in the final time segment.  
In order to observe long time behaviour of the two point function, we will expand in the powers of $e^{-ct}$. This regime of time is given by $\s{(E_2/2)}\ t \gg 1$ where $E_2>0$ (see blue part in Fig \ref{t-reg}), this gives us following expression
\begin{equation}\label{GE}
 G(t+ t_{12}, t ) = C_\Delta \left(\frac{c}{\sinh(ct_{12})}\right)^{2\Delta} \left[1 -\left(\frac{J \Delta \epsilon}{3b}\ \left[5\left( 1 +  \coth(c t_{12})\right) + 3 \sinh(c t_{12})\right]\right)\ e^{-c t} + O(e^{-2ct})\right].
\end{equation}
 
\subsubsection*{$E_2\to 0$ case} 

In the second real time segment our solutions have a special behaviour (\eq{fp0} and \eq{f0}), when 
\begin{equation}
 \left|\s{\f{E_2}{2}}\right|\ t \ll 1 \qquad \text{with }x_0=0. 
\end{equation}
We would like to see how does the 2 point function behave in this regime. The complete expression for this is messy but if we consider an additional  restriction on the time regime (see green part of Fig \ref{t-reg} ), 
\begin{equation}
  \left(\f{J \hat{\ep}}{4}\right)\ t \gg 1
\end{equation}
then we get  a simplified expression which is as follows
\begin{equation}\label{G0}
 G_{E_2 \to 0}(t + t_{12}, t) = \f{1}{\left(\f{J \hat{\ep}}{4}\ t_{12}\right)^{2\Delta}} 
 \left[ 1 - \left(\f{2\Delta\ t^2_{12}}{3}\right)\ \f{1}{t^2}   + \left(\f{2 \Delta\ 
  t^3_{12}}{3}\right) \f{1}{t^3} + O(t^{-4}) \right].
\end{equation}
\begin{figure}[H]
\centering
 \begin{tikzpicture}
 \draw [very thick,->](0,0)--(15,0);
 \draw [very thick,green] (3.5,0)--(5.5,0);
 \node [above] at (1.5,0) {$\f{4}{J \hat{\ep}}$};
 \draw (1.5,.1)--(1.5,-.1);
 \draw (7.5,.1)--(7.5,-.1);
 \node [above] at (7.5,0) {$\left|\s{\f{2}{E_2}}\right|$};
 \draw [very thick,orange] (9.5,0)--(14.9,0);
 \draw [->](7,-.5)--(8,-.5);
 \node  at (8.5,-.5) {\it{time}};
\end{tikzpicture}
\caption{Approximate regions of validity of of \eq{GE}(orange) and \eq{G0}(green) on the time axis.}
\label{t-reg}
\end{figure}
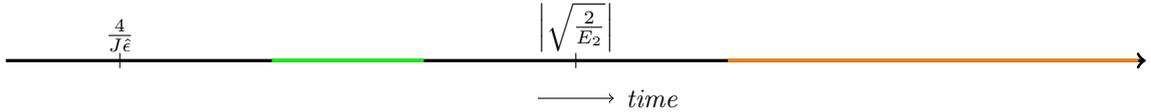
Intuitively when $E_2>0$ and $E_2 \to 0$ we are taking zero temperature limit of the thermal 2 point function and when $E_2<0$ the general structure of 2 point function is a power law behaviour plus oscillations but when $E_2 \to 0$ then the period of the oscillation becomes very large and the oscillations are no longer visible.

\section{Two coupled SYK models\label{mal-qi}}
In this section we work with two coupled SYK Hamiltonians with a coupling term, which was previously introduced in \cite{Maldacena:2018lmt}, 
\begin{equation}
 H = H_0 + H_{\text{int}} = H^L_{\text{syk}} + H^R_{\text{syk}} + H_{\text{int}}, \qquad H_{\text{int}} = \mu \left(i \sum_j \psi^L_j \psi^R_j\right). 
\end{equation}
We would like to look at dynamics of the soft modes in this model. In the soft sector the dynamical variables of the action are $t_l(u)$ and $t_r(u)$. The low energy effective action
\begin{equation}
 S = N \int du \left[-\frac{\alpha_s}{2}\left(\left\{\tan\frac{t_l(u)}{2},u \right\} + \left\{\tan\frac{t_r(u)}{2},u \right\}\right) + \mu \frac{c_\Delta}{(2\mathcal{J})^{2\Delta}} \left( \frac{t_l'(u)t'_r(u)}{\cos^2\frac{t_l(u)-t_r(u)}{2}}\right)^{\Delta} \right].
\end{equation}
Let us simplify this action by following re-definitions. (see \cite{Maldacena:2018lmt} for more details)
\begin{equation}
 \tilde u = \left(\f{\mathcal{J}}{\alpha_s}\right)\ u, \quad \eta = \left(\f{\mu \alpha_s}{\mathcal{J}}\right)\f{C_\Delta}{(2\alpha_s)^{2\Delta}}
\end{equation}
\begin{equation}\label{2seff}
 S = N \int d \tilde u \left[-\left(\left\{\tan\frac{t_l(\tilde u)}{2},\tilde u \right\} + \left\{\tan\frac{t_r(\tilde u)}{2},\tilde u \right\}\right) + \eta  \left(\frac{t'_l(u)t'_r(u)}{\cos^2\frac{t_l(u)-t_r(u)}{2}}\right)^{\Delta} \right]
\end{equation}
\subsection*{Global SL(2) symmetry}
The action \eq{2seff} has global SL(2) symmetry and the infinitesimal transformations on the fields are as follows
\begin{equation}
 \delta t_l = \epsilon^0 + \epsilon^+ e^{i t_l} + \epsilon^- e^{-i t_l}, \quad \delta t_r = \epsilon^0 - \epsilon^+ e^{i t_r} - \epsilon^- e^{-i t_r}.
\end{equation}
Let us assume that we have obtained a classical solution of the action \eq{2seff} such that $t_l(\tilde u) \neq t_r(\tilde u)$. Then it is always possible to make an SL(2) transformations to set $t_l(\tilde u) = t_r(\tilde u) = t(\tilde u)$ (see \cite{Maldacena:2018lmt} for more details). Then we can see that in this gauge condition, $Q_\pm =0$ for this solution. With the redefinition $t'(\tilde u) = e^{\tilde \phi(\tilde u)}$, we get 
\begin{equation}
 Q_0/N = 2 e^{-\tilde{\phi}} \left(- \tilde \phi'' - e^{2\tilde \phi} + \eta \Delta e^{2 \Delta \tilde \phi}\right) .
\end{equation}
$Q_0=0$ gives us
\begin{equation}\label{Q00}
 \tilde \phi'' = \eta \Delta e^{2 \Delta \tilde \phi} - e^{2\tilde \phi}. 
\end{equation}

\subsection*{Large N classical solutions}
Recall that, we can always get solutions of the form $t_l(\tilde u) = t_r(\tilde u) = t(\tilde u)$, hence instead of trying to get equations of motion from action \eq{2seff}, we try to get equations of motion for the following action
\begin{equation}
 S' = N \int d \tilde u \left[-2\left\{\tan\frac{t_l(\tilde u)}{2},\tilde u \right\}  + \eta  \left(t'_l(u)\right)^{2 \Delta} \right]. 
\end{equation}
Let us introduce a Lagrange multiplier $\lambda(\tilde u)$ which will set $t' = e^{\tilde \phi}$, then the action is 
\begin{equation}
 S' = N \int d\tilde u \ [\tilde \phi'^2 - e^{2\tilde \phi} + \eta e^{2\Delta \tilde \phi} + \lambda(\tilde u) \left(e^{\tilde \phi} - t'\right)]. 
\end{equation}
The equation of motions are 
\begin{align}\label{2sid-eom}
 \tilde \phi''(\tilde u) &= \f{\lambda}{2} e^{\tilde \phi} + \eta \Delta e^{2\Delta \tilde \phi} - e^{2\tilde \phi} = -\f{d \tilde V(\tilde \phi)}{d\tilde \phi} \nn \\
 \lambda'(\tilde u) &= 0, \nn \\
 t'(\tilde u) &= e^{\tilde \phi(\tilde u)},
\end{align}
following \eq{Q00} and \eq{2sid-eom}, $\lambda = 0$. With this input, \eq{2sid-eom} are very similar to \eq{phieom} and \eq{rest-eom} and we can exactly solve this problem for $\Delta = 1/2$ the expressions for potential and energy are 
\begin{equation}
 \tilde V(\tilde \phi) = \f{e^{2\tilde \phi}}{2} - \f{\eta }{2}\ e^{\tilde \phi}, \quad \tilde E = \f{\tilde \phi'^2}{2} + \f{e^{2\tilde \phi}}{2} - \f{\eta}{2} e^{\tilde \phi}
\end{equation}
\begin{figure}[H]
\centering
 \includegraphics[scale =.65]{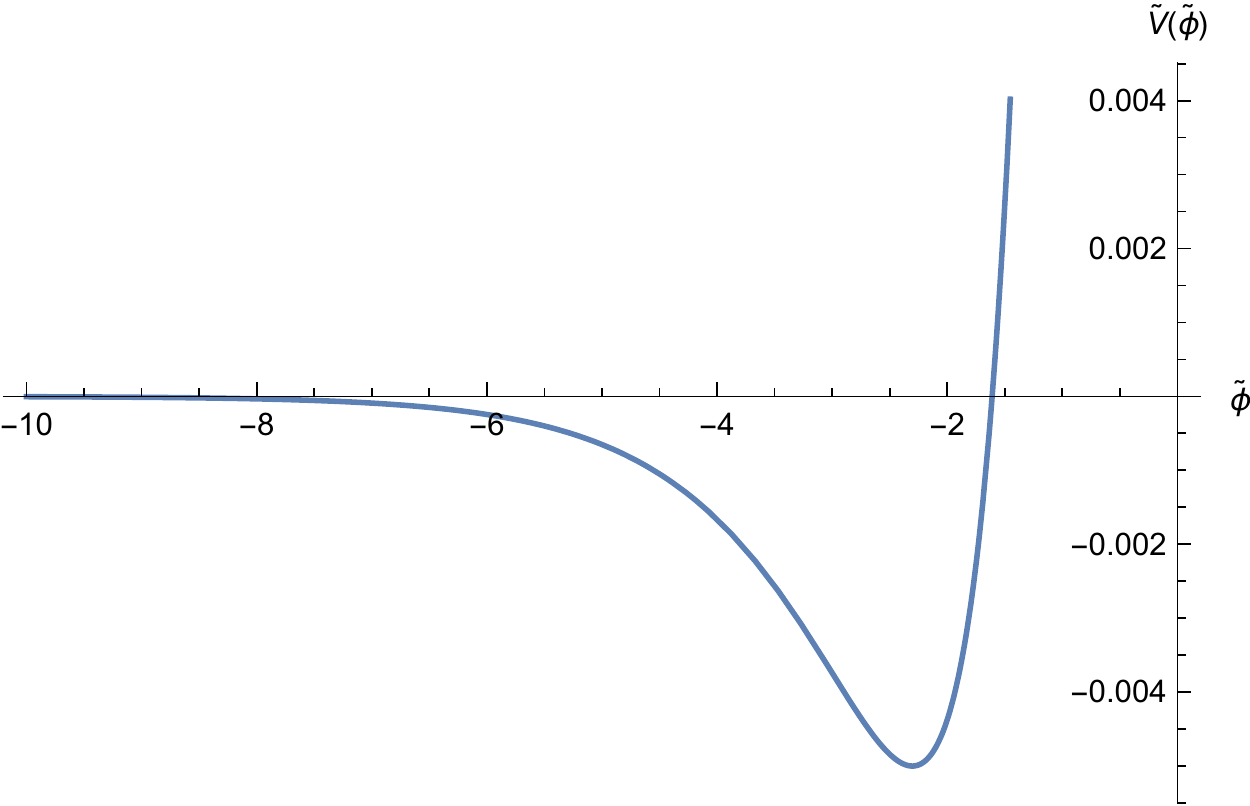}
 \caption{Plot $\tilde V(\tilde \phi)$ for $\eta = .2$}
\end{figure}
We can see that this problem also has two classes of solutions depending upon whether $\tilde E$ is positive or negative. As we saw in \eq{fp}, \eq{fbh} for the case of the modified SYK Hamiltonian \eq{eq:ham}. These solutions are given by   
\begin{equation}\label{2sid-phi}
 \tilde \phi(\tilde u) = -\log(\f{1}{4\tilde E}\ \left[\s{8\tilde E + \eta^2}\  \cos(\s{-2\tilde E}\ (\tilde u-\tilde u_0) + \theta)-\eta\right]).
\end{equation}
This solution is bounded for $\tilde E <0$ and unbounded for $\tilde E>0$. We can also design a quench protocol with $\eta$ as the quench parameter such that we start with $\tilde E <0$ and $\eta = \eta_1$ at $\tilde u = 0$ and then change this parameter to $\eta = \eta_2$ at some future time $\tilde u = T$ such that $\tilde E >0$. 

\par Let us start with $\tilde E_1 = -\eta_1^2/8$ which will make $\tilde \phi_1(\tilde u) = \log(\eta_1/2)$ (we choose $\eta_1$ according the boundary condition we would like to ensure on $\tilde \phi(\tilde u)$ at $\tilde u =0$) and $\tilde \phi'(\tilde u) = 0$, then energy post quench ($\tilde u > T $) is $\tilde E_2 = -(\eta_1 \eta_2/4)+(\eta_1)^2/8$, where $\eta_2$ is the changed value of the quench parameter $\eta$ from $\eta_1$. If the sign of energy $\tilde E$ is changed then \eq{2sid-phi} goes from being a bounded oscillatory function to an unbounded function or vice versa. As we have chosen $\tilde E_2 <0$ for which we have $\tilde \phi_1(\tilde u) = \log(\eta_1/2)$ this will be understood as an periodic solution with any specified period. With this understanding, if  $\eta_2 < \eta_{cr}$ where $\eta_{cr} = \eta_1/2$ we will have $\tilde E_2>0$ and we will get following unbounded solution.  
\begin{align}
  \tilde \phi_2(\tilde u) &= -\log(\f{1}{4\tilde E_2}\ \left[\s{8\tilde E_2 + \eta_2^2}\  \cosh(\s{2\tilde E}\ (\tilde u-\tilde u_0) + \theta)-\eta_2\right]),
  \nonumber\\ \tilde E_2 &= \f{\eta_1^2}{8}-\left(\f{\eta_1 \eta_2}{4}\right)
  =  \f{\eta_1}{4}(\eta_{cr}-\eta_2), \; \eta_{cr}=\eta_1/2
  \label{phi-2sid}
\end{align}
The situation here is similar to what is described in
\autoref{sec-choptuik}. Once again, there is a range of the perturbation
$\eta_2$, viz.
$\eta_2<\eta_{cr}$ for which no black hole is formed. On the other
hand, for $\eta_2> \eta_{cr}$ a black hole is formed.
The temperature of the black hole can be determined by looking
at the imaginary time period of \eq{phi-2sid}, at large times, as in
\autoref{sec-choptuik}. We find that 
\begin{align}\label{2sid-T}
 T_{bh} = \f{1}{\pi} \s{\f{\tilde E_2}{2}} = \left(\f{e^{-\tilde \phi(T)/2}}{\pi\s{2}}\right)\left(\f{\tilde \phi'(T) + e^{2\tilde \phi(T)}}{e^{\tilde \phi}} - \eta_2 \right)^{1/2} = A (\eta_{cr} - \eta_2)^{1/2}
\end{align}

\section{Discussion and open questions}

\begin{itemize}
 \item In this paper, we have discussed a solvable model of gravitational
collapse, namely a deformed time dependent SYK model. We performed a sudden quench
in the deformation parameter, which was represented by a sudden jump of the
bulk stress tensor at the corresponding time. We showed that there is
a critical value of the deformation parameter, which demarcates between (a)
set of initial conditions from which gravitational collapse leads to a
black hole, and (b) the set of initial conditions which leads to a
horizonless geometry. This is akin to Choptuik criticality in
gravitational collapse. In case the black hole does form, we found that the black hole temperature is given by a power low 
$T_{bh} \propto (\Delta \ep - \Delta \ep_{cr})^{1/2}$.

\item Our discussion of gravitational collapse showed a universality. Starting with the ground state of an SYK model, deformed by an operator $\ep H_M^{(s)}$, if we perform a sudden quench $\ep \to 0$, the resulting black hole does not have the memory of `$s$'. It would be interesting to count the number of possible initial states described as above and see if can understand the Bekenstein-Hawking entropy of the resulting black hole from such a calculation.

\item We would like to understand the transition to chaotic behaviour as we cross over to the parameter region corresponding to black holes. Such a transition can be verified by a computation of the Lyapunov index through out of time ordered correlators. Work in this direction is in progress.

\item We would also like to better understand the nature of criticality in the $\ep$ parameter from the point of view of the boundary theory. In particular we would like to understand why, 
even though it is a strongly coupled theory, the two point function shows a periodic behaviour instead of thermalizing, for $\Delta \ep < \Delta \ep_{cr}$. 

\item It would be interesting to better understand the bulk dual corresponding to Kourkoulou-Maldacena perturbation. Especially in view of the fact that it is not an $O(N)$ invariant operator.\footnote{It is useful to contrast with the two sided case, discussed in section \ref{mal-qi}, where the perturbation operator is $O(N)$ invariant}
\end{itemize}

\subsection*{Acknowledgment}

We would like to thank Dionysios Anninos, Micha Berkooz, Shiraz
Minwalla, Kyriakos Papadodimas, Mukund Rangamani, Suvrat Raju, Subir
Sachdev, Douglas Stanford, Andy Strominger and especially Juan
Maldacena for illuminating discussions. The work of A.G, L.K.J
and G.M was supported in part by the Infosys foundation for the study
of quantum structure of spacetime. The work of A.G and G.M was
supported in part by the International Centre for Theoretical Sciences
(ICTS) during various visits, in particular for the program -
``AdS/CFT at 20 and Beyond''. S.R.W would like to acknowledge CERN
Theory Division where part of this work was done, and the Infosys
Foundation Homi Bhabha Chair for their generous support. G.M and S.R.W
would like to thank KITP, Santa Barbara, for the stimulating program
``Order from Chaos'' where several important aspects of this work were
discussed at the finishing stage. This research was supported in part
by the National Science Foundation Grants No. 1748958 and the Simons
Distinguished Visitors Program at KITP.

\appendix

\section{Details of the derivation of the action \eq{full-action}}\label{app-path-integral}
The aim of this section is to derive the form of the effective action at low energies governing the path integral at $t>0$. In order to identify the action we will consider following two point correlation function  
\begin{equation}\label{2pt}
  G(t,t',l,l') = \f{\braket{B_s(l)|\hat \psi_a(t) \hat \psi_a(t')|B_s(l')}}{\braket{B_s(l)|B_s(l')}}.
\end{equation}
Let us write above quantity in the interaction picture, where 
\begin{equation}
 H = H_0 + \ep(t) \ H^{(s')}_M,\quad \text{ with } H_M^{(s')} = \left(-\f{ J}{2}\right) \sum_{k=1}^{N/2} s'_k \hat S_k \quad \text{ and } H^{(s')}_I(t)= e^{i H_0 t}\ H^{s'}_M\ e^{-i H_0 t}.
\end{equation}
We assume that $\ep(t)=0$ for $t<0$ i.e. the interaction is turned on at $t=0$.
Note that the $k$'th component of the spin vector $s$ in the $\ket{B_s(l)}$ is different from the number $s'_k$ which appear in $H^{(s')}_M$ and $H_0$ is given in \eq{eq:ham} (hereafter we will omit the superscript of $H^{(s')}_M$).

Let us also define following evolution operators,
\begin{equation}
 U(t,0)  = \mathcal{T}\left[\exp\left(-i \int_{0}^t dt''  H(t'')\right)\right], \label{U}\hspace{9cm} \\
 V(t,0)=e^{iH_0 t}\ U(t,0) =\mathcal{T}\left[\exp\left(-i \int_{0}^t dt'' \ep(t'') H_I(t'')\right)\right] \hspace{7cm}\nn \\
 \hspace{-.2cm}= \sum_{n=0}^{\infty} \left(\f{(-i)^n}{n!}\right) \int_0^t dt_1 \cdots \int^t_0 dt_n\ \left(\prod_{i=1}^n \ep(t_i)\right) \mathcal{T} \left[H_I(t_1) \cdots H_I(t_n) \right]\nn \\
 \hspace{2.8cm}= \sum_{n,k_i} \left(\f{iJ\epsilon}{2}\right)^n\ \f{1}{n!} \int_0^t dt_1 \cdots \int^t_0 dt_n\ \left(\prod_{i=1}^n \ep(t_i)\right)\mathcal{T} \left[\hat{S}_{k_1}(t_1) \cdots \hat{S}_{k_n}(t_n) \right]s'_{k_1}\cdots s'_{k_n}\label{V}.
\end{equation}
Using the above definitions we can write an operator in the Heisenberg picture as
\begin{align}\label{ctp-1}
 \mathcal{O}(t') &= U^{\dagger}(t',0)\ \mathcal{O}(0)\ U(t',0)\nn \\ &= V^{\dagger}(t',0)\ \mathcal{O}_I(t')\ V(t',0) \nn \\ &= V^{\dagger}(t',0)\ V(t',\infty)V(\infty,t')\ \mathcal{O}_I(t')\ V(t',0). 
\end{align}
Here the subscript of $\mathcal{O}_I(t')$ indicates that it is an interaction picture operator and we have inserted an identity $V(t',\infty)V(\infty,t') =1$ on the left of $O_I(t')$. $\mathcal{O}(t')$ given by \eq{ctp-1} can be equivalently written as 
\begin{equation}\label{ctp-2}
 \mathcal{O}(t') = \mathcal{T}_{ctp} \left[\exp\left(-i \int_{ctp} dz \ \ep(z)\ H_I(z)\right)\mathcal{O}_I(t')\right],
\end{equation}
where $\mathcal{T}_{ctp}$ is the time ordering defined on a ``closed time path" (CTP) drawn in \autoref{CTP}. We also define $z$ as variable along the CTP (the equivalence of \eq{ctp-1} and \eq{ctp-2} can be shown by simply expanding time ordered exponential in both expressions). 
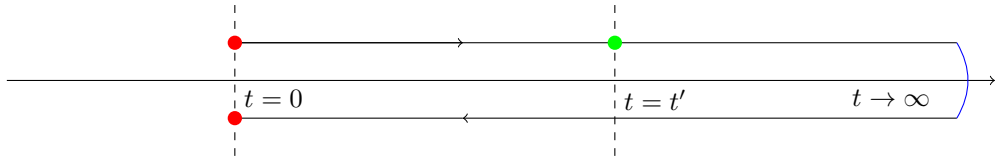
\begin{figure}[H]
 \centering
 \begin{tikzpicture}
  \draw [->](0,0)--(13,0);
  \draw [->](3,.5)--(6,.5);
  \draw (3,.5)--(12.5,.5);
  \draw [->](12.5,-.5)--(6,-.5);
  \draw (6,-.5)--(3,-.5);
  \draw [dashed] (3,1)--(3,-1);
  \node [below right] at (3,0) {$ t =0$};
  \node [below right] at (11,0) {$ t \to \infty$};
  \draw [dashed] (8,1)--(8,-1);
  \node [below right] at (8,0) {$t = t'$}; 
\draw[blue] (12.5,.5) to [out=300,in=60] (12.5,-.5);
  \draw[fill,red,thick] (3,.5) circle [radius=0.08];
  \draw[fill,red,thick] (3,-.5) circle [radius=0.08];
  \draw[fill,green,thick] (8,.5) circle [radius=0.08];
 \end{tikzpicture}
\caption{Closed time path: We start at $t=0$ with $\ket{B_s(l)}$ (upper red dot), then evolve up to $t=t'$ insert the operator $\mathcal{O}$ at $t=t'$, then go all the way up to $\infty$ and then back to $t=0$ along the lower time segment where we end with $\bra{B_s(l)}$ (lower red dot). Note that above, the upper and lower time segments are exactly along the real line, we have separated them in the figure only for visual clarity; Note that the blue curve is actually a single point}
\label{CTP}
\end{figure}
Now we use \eq{ctp-2} to write the numerator of \eq{2pt} which we denote as $\bar N$, as follows
\begin{equation}
\bar N =  {\Braket{B_s(l)|\mathcal{T}_{ctp} \left[\exp\left(-i \int_{ctp} dz\ \ep(z) \ H_I(z)\right)\hat{\psi}_{I,a}(t)\right]\ \mathcal{T}_{ctp} \left[\exp\left(-i \int_{ctp} dz\ \ep(z) \   H_I(z)\right)\hat{\psi}_{I,a}(t')\right]|B_s(l')}}.
\end{equation}
Assuming $t'<t$ we get
 \begin{equation}\label{nbar}
 \bar N =  {\Braket{B_s(l)|\mathcal{T}_{ctp} \left[\exp\left(-i \int_{ctp} dz\ \ep(z)\  H_I(z)\right)\hat{\psi}_{I,a}(t)\ \hat{\psi}_{I,a}(t')\right]|B_s(l')}}.
\end{equation}
The pictorial representation of the contour for the above expression is given in \autoref{ctp}.
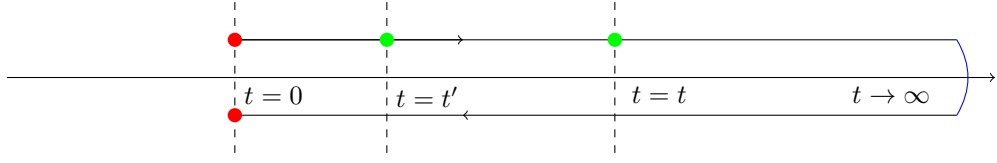
\begin{figure}[H]
\centering
 \begin{tikzpicture}
  \draw [->](0,0)--(13,0);
  \draw [->](3,.5)--(6,.5);
  \draw (3,.5)--(12.5,.5);
  \draw [->](12.5,-.5)--(6,-.5);
  \draw (6,-.5)--(3,-.5);
  \draw [dashed] (3,1)--(3,-1);
  \draw [dashed] (5,1)--(5,-1);
  \node [below right] at (3,0) {$ t =0$}; 
  \draw [dashed] (8,1)--(8,-1);
  \node [below right] at (8.1,0) {$t = t$};
  \node [below right] at (5,0) {$t = t'$};
  \node [below right] at (11,0) {$t \to \infty$};
  \draw[blue] (12.5,.5) to [out=300,in=60] (12.5,-.5);
  \draw[fill,red,thick] (3,.5) circle [radius=0.08];
  \draw[fill,red,thick] (3,-.5) circle [radius=0.08];
  \draw[fill,green,thick] (5,.5) circle [radius=0.08];
  \draw[fill,green,thick] (8,.5) circle [radius=0.08];
 \end{tikzpicture}
\caption{We have CTP described in \autoref{CTP} with two operator insertions $\psi_I(t)$ and $\psi_I(t')$, both along the upper time segment. The operator insertion points are indicated by the green dots on the contour}
 \label{ctp}
\end{figure}
We will now expand the time ordered exponential in $\bar N$, explicitly 
\begin{align}\label{def-A}
 & \bar{N} = \Braket{B_s(l)|\mathcal{T}_{ctp}\left[\exp\left(-i \int_{ctp} dz\ \ep(z) H_I(z)\right) \hat{\psi}_{I,a}(t)\ \hat{\psi}_{I,a}(t')\ \right]|B_s(l')} \nn \\
 &= \Braket{B_s(l)|\left[\sum_{n,k_i} \left(\f{(iJ/2)^n}{n!}\right) \int_{ctp} \left( \prod_{i=1}^n dz_i \ep(z_i)\right)\mathcal{T}_{ctp} \left[\hat{S}_{I,k_1}(z_1) \cdots \hat{S}_{I,k_n}(z_n)\ \hat{\psi}_{I,a}(t)\ \hat{\psi}_{I,a}(t') \right]s'_{k_1}\cdots s'_{k_n}\right]|B_s(l')} \nn \\
 &= \sum_{n,k_i} \left(\f{(iJ/2)^n}{n!}\right) \int_{ctp} \left( \prod_{i=1}^n dz_i \ep(z_i)\right)\ \Braket{B_s(l)|\mathcal{T}_{ctp} \left[\hat{S}_{I,k_1}(z_1) \cdots \hat{S}_{I,k_n}(z_n)\ \hat{\psi}_{I,a}(t)\ \hat{\psi}_{I,a}(t') \right]s'_{k_1}\cdots s'_{k_n}|B_s(l')} .
 \end{align}
All the operators in the above expressions are in the interaction picture. Hereafter we will omit the subscript 'I' and all the operators should be understood to be in the interaction picture. Now we will introduce identity operator $\hat{S}^2_{k_i} = 1$ in the above expression, as follows 
\begin{align}
& \bar{N} = \Braket{B_s(l)|\mathcal{T}_{ctp}\left[\exp\left(-i \int_{ctp} dz\ \ep(z)\ H_I(z)\right) \hat{\psi}_{I,a}(t)\ \hat{\psi}_{I,a}(t')\ \right]|B_s(l')} \nn \\
 &= \sum_{n,k_i} \left(\f{(iJ/2)^n}{n!}\right) \int_{ctp} \left( \prod_{i=1}^n dz_i \ep(z_i)\right)\Braket{B_s|e^{-lH_0}\ \mathcal{T}_{ctp} \left[\hat{S}_{k_1}(z_1) .. \hat{S}_{k_n}(z_n)\ \hat{\psi}_{a}(t)\ \hat{\psi}_{a}(t') \right]e^{-l'H_0}s'_{k_1}.. s'_{k_n}\left(\prod_{i=1}^n \hat{S}^2_{k_i}\right)|B_s} \nn \\
&= \sum_{n,k_i} \left(\f{(iJ/2)^n}{n!}\right) \int_{ctp} \left( \prod_{i=1}^n dz_i \ep(z_i)\right) \Braket{B_s|e^{-lH_0}\ \mathcal{T}_{ctp} \left[\hat{\psi}_{a}(t)\ \hat{\psi}_{a}(t')\ \prod_{i=1}^n  \hat{S}_{k_i}(z_i)\right] e^{-l'H_0} \left(\prod_{j=1}^n\hat{S}_{k_j}\ s'_{k_j} s_{k_j}\right) |B_s}.
\end{align}
Using this result we can write
\begin{equation}\label{fin}
G(t,t',l,l') = \sum_{n=0}^{\infty} \left(\f{1}{n!}\right)\left(\f{iJ}{2}\right)^n \int_{ctp} dz_1 \ep(z_1) \cdots \int_{ctp} dz_n\ep(z_n)\ \mathcal{A}^s_n[j_{abcd}]
\end{equation}
with
\begin{equation}\label{torder}
 \mathcal{A}^s_n[j_{abcd}]=\f{\Braket{B_s|e^{-lH_0}\ \mathcal{T}_{ctp} \left[\hat{\psi}_{a}(t)\ \hat{\psi}_{a}(t')\ \sum_{k_i=1}^{n}\prod_{i=1}^n  \hat{S}_{k_i}(z_i)\right] e^{-l'H_0}\left(\prod_{j=1}^n\hat{S}_{k_j}\ s'_{k_j} s_{k_j}\right) |B_s}}{\braket{B_s(l)|B_s(l')}}
\end{equation}
Note that the operators $\hat{S}_{k_i}$ do not have a time argument, it is inserted immediately after the state $\ket{B_s}$ and superscript in $\mathcal{A}^s_n[j_{abcd}]$ is to indicate the boundary conditions $\ket{B_s}$.

Now lets take a careful look at $\mathcal{A}_n$ reading from right to left, we will extend our path integral contour by adding two Euclidean paths before and after CTP (see \autoref{CTP}).Let us call the new contour `c' and it is given by \autoref{c}.

We will use the same notation for the variable along this contour except $z$ now takes values on the complex plane. In our convention we start at $z=il'$ (in this convention $z$ along CTP is real) with $\ket{B_s}$ state, then insert $n$ $\hat S$ operators (indicated by first green line in \autoref{c}) then we evolve in $-ve$ imaginary time for $l'$ distance and reach $z=0$ (the operator $e^{-l'H_0}$ evolves the state in the negative imaginary time direction), now we have $n+2$ operator insertions on the CTP, $2$ of these operators, $\psi$'s  have fixed positions, $t$ and $t'$ indicated by green lines in \autoref{c} and rest $n$ are the interaction Hamiltonians inserted at some arbitrary positions on the CTP (The positions don't really matter as we will be integrating over these positions keeping the time ordering along the contour). The CTP ends back at $z=0$ from there we evolve $l$ distance along $-ve$ imaginary time direction and reach $z=-il$ and take an inner product with $\bra{B_s}$ 
\begin{figure}[H]
 \centering
 \begin{tikzpicture}
 \node[above right] at (3,3.1) {$z = il'$};
  \node[below right] at (3,-2.1) {$z = -il$};
  \draw [->](3,-2.5)--(3,4);
  \draw [->,red,thick] (3,3)--(3,2);
  \draw [red,thick](3,2)--(3,0.5);
  \draw [->,red,thick] (3,-.5)--(3,-1);
  \draw [red,thick] (3,-1)--(3,-2);
  \draw [->](0,0)--(13,0);
  \draw [->,thick](3,.5)--(6,.5);
  \draw [thick](3,.5)--(12.5,.5);
  \draw [->,thick](12.5,-.5)--(6,-.5);
  \draw [thick](6,-.5)--(3,-.5);
  \draw [dashed] (5,1)--(5,-1);
  \draw[fill,black,thick] (3,3) circle [radius=0.08];
  \draw[fill,black,thick] (3,-2) circle [radius=0.08];
  \draw[fill,green,thick] (3,2.9) circle [radius=0.08];
  \draw[fill,green,thick] (5,.5) circle [radius=0.08];

  \node [below right] at (3,0) {$ t =0$}; 
  \draw [dashed] (8,1)--(8,-1);
  \node [below right] at (8.1,0) {$t = t$};
  \node [below right] at (5,0) {$t = t'$};
  \node [below right] at (11,0) {$t \to \infty$};
  \draw[blue] (12.5,.5) to [out=300,in=60] (12.5,-.5);
  \draw[blue,very thick,dashed] (0,3)--(10,3);
  \draw[blue,very thick,dashed] (0,-2)--(10,-2);
  \draw[fill,green,thick] (8,.5) circle [radius=0.08];
  \draw[decoration={brace,mirror,raise=5pt},decorate] (3,.6) -- node[right=8pt] {$l'$} (3,3);
  \draw[decoration={brace,raise=5pt},decorate] (3,-.6) -- node[right=8pt] {$l$} (3,-2);
 \end{tikzpicture}
\caption{ Contour `c': the black dots are the states $\ket{B_s}$ (above) and $\bra{B_s}$ (below). If we have insertions of only flip symmetric operators then the boundary conditions (black dots) can be converted to trace boundary condition, in this case we identify the two blue lines and convert the contour from a segment to a circle}
\label{c}
\end{figure}
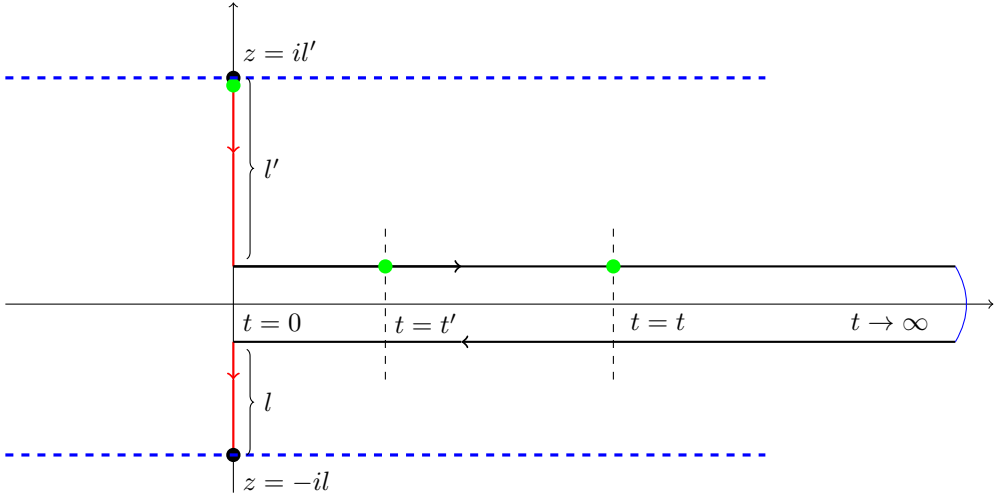
\subsubsection*{Flip transformation}
The expectation value of product of time ordered operators in \eq{torder} can be evaluated by a path integral 
\begin{equation}\label{pintA}
 \mathcal{A}^s_n[j_{abcd}] = \f{\int \boldsymbol{\mathcal{D}\psi}\ \braket{B_s|\psi^{-l'}}\ \braket{\psi^{l}|B_s}\ e^{-S^{E_1}_{syk}[\psi]}\  
 \left[\sum_{k_i}\ \prod_{i}  \hat{S}_{k_i}(\tau_i)\hat{S}_{k_i}(l')\ \psi_{a}(\tau_1)\ \psi_{a}(\tu_2)\left(s'_{k_i} s_{k_i}\right)\right]}{\int \boldsymbol{\mathcal{D}\psi}\ \braket{B_s|\psi^{-l'}}\ \braket{\psi^{l}|B_s}\ e^{-S^{E_1}_{syk}[\psi]}},
\end{equation}
with
\begin{equation}\label{SE1}
 S^{E_1}_{syk}[\psi] = \f{1}{2}\int d\tu\  \psi_i(\tu)\del_\tu\psi_i(\tu) + \sum_{i<j<k<l} \int d\tu\  j_{ijkl}\ \psi_i(\tu) \psi_j(\tu) \psi_k(\tu) \psi_l(\tu).
\end{equation}
where $\hat\psi_i \ket{\psi} = \psi_i \ket{\psi}$ and $\boldsymbol{\mathcal{D}\psi}$ is the appropriate measure. Note that we have written a path integral looks completely Euclidean even though the contour contains CTP. In order to compute \eq{torder}, we first obtain answer for a Euclidean path integral along a line segment of length $l+l'$ and $\ket{B_s}$ as boundary conditions with insertions along CTP replaced by insertions at some proxy Euclidean time ($\tau_1,\tau_2\text{ and }\tau_i$'s), we then analytically continue these proxy Euclidean times in the obtained answer to Lorentzian signature ($it,it'\text{ and }z_i$'s) so that they are appropriately situated on CTP. Note that we are using the variable $\tau = -\text{Im}[z] $. 

In order to write the path integral we divide the path in Fig \ref{c} in $M$ time steps which gives us $M+1$ time slices and insert identities between adjacent slices. We have denoted identity inserted between $0$th and the $1$st slice as  $\int \tilde d \psi^{-l'} \ket{\psi^{-l'}}\bra{\psi^{-l'}}$ ( $\tilde d \psi^{-l'}$ is an appropriate measure) and the one inserted between last two slices (labeled by $M$ and $M-1$) as $\int \tilde d \psi^{l} \ket{\psi^{l}}\bra{\psi^{l}}$. The measure of all of these identities contribute to the kinetic term in the Lagragian except $\tilde d \psi^{l}$ hence formally 
\begin{equation}
 \boldsymbol{\mathcal{D}\psi} =\tilde d \psi^{l}\ \prod_{i=1}^{M} d\psi^{i} 
\end{equation}

The algebra of $N$ Majorana fermions coincides with $SO(N)$ Clifford algebra. The Flip operator in the $2^{N/2}$ dimensional Hilbert space spanned by $\ket{B_s}$ is given by $F_k = \s{2} \hat{\psi}_{2k-1}$, flips the spin on the $k$'th site. The action on the fermions is given as follows
\begin{equation}
  F_k^2=1, \qquad \{F_k,\hat{\psi}_{2k}\}=0, \qquad [F_k,\hat{\psi}_{j}]=0\ \forall\ j\neq2k.
\end{equation}
A generic flip operator would be product of flip operators on different sites
\begin{equation}
 F = F_{k_1}\cdots F_{k_m}, \qquad m < N/2
\end{equation}
Now we will introduce these operators in \eq{torder}
\begin{equation}
  \mathcal{A}^s_n[j_{abcd}]=\f{\Braket{B_s|(F^2)e^{-lH_0}\ \mathcal{T}_{ctp} \left[\hat{\psi}_{a}(t)\ \hat{\psi}_{a}(t')\ \sum_{k_i=1}^{n}\prod_{i=1}^n  \hat{S}_{k_i}(z_i)\right] e^{-l'H_0}\left(\prod_{j=1}^n\hat{S}_{k_j}\ s'_{k_j} s_{k_j}\right) (F^2)|B_s}}{\braket{B_s|(F^2)e^{-2lH_0}(F^2)|B_s}}
\end{equation}
in the the above expression $H_0$ is defined in \eq{eq:ham0}, then we have 
\begin{equation}
 F H_0 F = \sum_{a<b<c<d} j_{abcd}\ F \hat\psi_a \hat\psi_b \hat\psi_c \hat\psi_d F.
\end{equation}
In the above expression, odd fermions remain the same, but if $F$ is acting on an even fermion then it picks a $-ve$ sign. We can absorb these $-ve$ signs in $j_{abcd}$ and define
\begin{equation}
 \tilde H_0 = F H_0 F = \sum_{a<b<c<d} \tilde j_{abcd}\ \hat\psi_a \hat\psi_b \hat\psi_c \hat\psi_d.
\end{equation}
Then we get
\begin{equation}\label{ftorder}
  \mathcal{A}^s_n[j_{abcd}]=\f{\Braket{B_{s'}|\exp(-l\tilde{H}_0)\ \mathcal{T}_{ctp} \left[\hat{\psi}_{a}(t)\ \hat{\psi}_{a}(t')\ \sum_{k_i=1}^{n}\prod_{i=1}^n  \hat{S}_{k_i}(z_i)\right] \exp(-l'\tilde{H}_0)\left(\prod_{j=1}^n\hat{S}_{k_j}\ s'_{k_j} s_{k_j}\right) |B_{s'}}}{\braket{B_{s'}|\exp(-2l\tilde{H}_0)|B_{s'}}}
\end{equation}
where $F\ket{B_s} = \ket{B_s'}$ and the evolution in \eq{ftorder} is done with $\tilde H_0$. Note that signs picked up by all the other insertions cancel out. Now we can also write \eq{ftorder} as  
\begin{equation}\label{fpintA}
 \mathcal{A}^{s'}_n[\tilde j_{abcd}] = \f{\int \boldsymbol{\mathcal{D}\psi}\ \braket{B_{s'}|\psi^{-l'}}\ \braket{\psi^{l}|B_{s'}}\ e^{-\tilde{S}^{E_1}_{syk}[\psi]}\  
 \left[\sum_{k_i}\ \prod_{i}  \hat{S}_{k_i}(\tu_i)\hat{S}_{k_i}(l')\ \psi_{a}(\tu_1)\ \psi_{a}(\tu_2)\left(s'_{k_i} s_{k_i}\right)\right]}{\int \boldsymbol{\mathcal{D}\psi} \braket{B_{s'}|\psi^{-l'}}\ \braket{\psi^{l}|B_{s'}}\ e^{-\tilde{S}^{E_1}_{syk}[\psi]}},
\end{equation}
with
\begin{equation}\label{SE11}
 \tilde{S}^{E_1}_{syk}[\psi] = \f{1}{2}\int d\tu\  \psi_i(\tu)\del_\tu\psi_i(\tu) + \sum_{i<j<k<l} \int d\tu\  \tilde j_{ijkl}\ \psi_i(\tu) \psi_j(\tu) \psi_k(\tu) \psi_l(\tu).
\end{equation}
As all we have done is to introduce identities in $\eq{pintA}$, $\mathcal{A}^s_n[j_{abcd}] = \mathcal{A}^{s'}_n[\tilde j_{abcd}]$.
\subsubsection*{Disorder averaging at large N}
We are dealing with a disordered system \eq{SE1}, hence observable $\mathcal{O}$ depends on the disorder $j_{abcd}$ and we should do a quenched averaging of the disorder 
\begin{equation}
 \tilde{\mathcal{O}} = \int \prod_{a<b<c<b} \f{dj_{abcd}}{\s{2\pi \sigma^2}}\ \exp\left(-\f{j^2_{abcd}}{2\sigma^2}\right)\ \mathcal{O}, \qquad \text{with } \sigma^2 = \f{3! J^2}{N^3}
\end{equation}
In SYK model at large $N$ answer can be obtained by annealed averaging, where we integrate the disorder from numerator and the denominator separately.  Generally
\begin{equation}
 \int \prod_{a<b<c<b} \f{dj_{abcd}}{\s{2\pi \sigma^2}}\ \exp\left(-\f{j^2_{abcd}}{2\sigma^2}\right)\ \left[\int \mathcal{D}\psi\ e^{-S^{E_1}_{syk}}\ (\cdots)\right] \propto \int \mathcal{D}\psi\ e^{-S^{E_2}_{syk}}\ (\cdots). 
\end{equation}
Here $(\cdots)$ indicates disorder independent product of operators and
\begin{equation}\label{SE2}
 S^{E_2}_{syk} = \f{1}{2}\int d\tu\  \psi_i(\tu)\del_\tu\psi_i(\tu) - \f{NJ^2}{8}\ \iint d\tu d\tu'\ \left(\f{1}{N} \sum_i \psi_i(\tu)\psi_i(\tu')\right)^4
\end{equation}
We can see that the action \eq{SE2} does not have any indication of the flip transformation unlike \eq{SE11} as all it did was to change the disorder $j_{abcd}$. Let 
\begin{equation}
 \tilde{\mathcal{A}}^s_n = \int \prod_{a<b<c<b} \f{dj_{abcd}}{\s{2\pi \sigma^2}}\ \exp\left(-\f{j^2_{abcd}}{2\sigma^2}\right)\ \mathcal{A}^s_n[j_{abcd}]
\end{equation}
then the only difference between $\tilde{\mathcal{A}}^s_n$ and $\tilde{\mathcal{A}}^{s'}_n$  is in the boundary conditions ($\ket{B_s}$ changes to $\ket{B_{s'}}$). The averaging is done separately in numerator and denominator, hence the fact that flip insertion only affect boundary conditions is separately true for numerator and denominator of $\tilde{\mathcal{A}}^s_n$. 

Let $\tilde{N}^s_n$ and $\tilde{D}^s$ be annealed averages of the numerator and denominator of \eq{pintA}, then $\tilde{N}^s_n = \tilde{N}^{s'}_n$ and $\tilde{D}^s = \tilde{D}^{s'}$. Remember that we have not restriction on the choice of $F$, hence we could sum over the complete group and divide by the volume ($2^{N/2}$). This allows us to do following replacement
\begin{equation}
 \braket{B_{s'}|\psi_{-il}}\ \braket{\psi_{il'}|B_{s'}} = 2^{-N/2}\ \text{Tr}\left(\ \ket{\psi_{-il}} \bra{\psi_{il'}}\ \right)
\end{equation}
The factor of $2^{-N/2}$ from numerator will cancel with the same factor from denominator. The dependence on the boundary conditions completely vanishes with this replacement and we get
\begin{equation}
 \tilde{\mathcal{A}}_n = \f{\oint \boldsymbol{\mathcal{D}\psi}\ e^{-S^{E_2}_{syk}[\psi]}\  
 \left[\sum_{k_i}\ \prod_{i}  \hat{S}_{k_i}(\tu_i)\hat{S}_{k_i}(l')\ \psi_{a}(\tu_1)\ \psi_{a}(\tu_2)\left(s'_{k_i} s_{k_i}\right)\right]}{\oint \boldsymbol{\mathcal{D}\psi}\ e^{-S^{E_2}_{syk}[\psi]}}.
\end{equation}
Tracing puts our theory on a Euclidean circle of length $l+l'$ (this implies identifying two blue lines in \autoref{c}) and anti-periodic boundary of conditions on fermionic fields. 

We can introduce a Lagrange multiplier bilocal field $\Sigma$ in the path integral to set the combination $1/N\sum \psi_i\psi_i$ equal to another bilocal field $G$ and then integrate out the fermions from path integral. We get a path integral over $\Sigma, G$ variables with following action 
\begin{equation}\label{SE3}
 S^{E_3}_{\text{syk}}[G,\Sigma] = -\f{N}{2}\log \text{det}(\del_\tu-\Sigma) + \f{N}{2} \int d\tu d\tu' \left[ \Sigma(\tu,\tu')G(\tu,\tu') - \f{ J^2}{4} G^4(\tu,\tu')\right].
\end{equation}
With these variables
\begin{equation}
 \tilde{\mathcal{A}}_n = \f{\oint \mathcal{D}G\ \mathcal{D}\Sigma\ e^{-S^{E_3}_{syk}[G,\Sigma]}\  
 \left[\sum_{k_i}\ \prod_{i}  \hat{S}_{k_i}(\tu_i)\hat{S}_{k_i}(l')\ \psi_{a}(\tu_1)\ \psi_{a}(\tu_2)\left(s'_{k_i} s_{k_i}\right)\right]}{\oint \mathcal{D}G\ \mathcal{D}\Sigma\ e^{-S^{E_3}_{syk}[G,\Sigma]}}.
 \label{zn}
\end{equation}
The notation $\oint$ is to indicate that now we have trace boundary conditions, which is identifying two blue line in \autoref{c}.

As explained earlier the computation of \eq{torder} contains path integral along CTP where we have to analytically continue the action to Lorentzian signature. Alternatively \eq{torder} can be computed from Euclidean path integral, \eq{zn}, over a circle of length $l+l'$ with operator insertions on the Euclidean time circle and then appropriately analytically continuing operator positions to real time in the answer. In order to take this approach we should start with
\begin{equation}\label{znp}
 \bar{\mathcal{A}}_n =\f{1}{\mathcal{Z}(l+l')} \oint \mathcal{D}G\ \mathcal{D}\Sigma\ \ e^{-S^{E_3}_{syk}[G,\Sigma]}\  
 \left[\sum_{k_i}\ \prod_{i}  \hat{S}_{k_i}(\tu_i)\hat{S}_{k_i}(-l')\ \psi_{a}(\tu_1)\ \psi_{a}(\tu_2)\left(s'_{k_i} s_{k_i}\right)\right]
\end{equation}
 all the operator insertions above are along Euclidean time and at the end of the calculation we will analytically continue $\tu_1=it$, $\tu_2=it'$ and $\tu_i = i t_i$. 
 
 \subsubsection*{Saddle point approximation, conformal symmetry and soft modes}

A path integral involving action \eq{SE3} can be approximated by it saddle point at large $N$. The classical equations of motions are 
\begin{equation}
 \f{1}{G(w)} = -iw -\Sigma(w), \qquad \Sigma(\tu,\tu') = J^2\ G(\tu,\tu')^{3}
\end{equation}
at low energies $J\gg w$ we drop the $w$ term in the first equation as $\Sigma$ scales with $J^2$. This is also equivalent to dropping the $\del_{\tu}$ term from the action. In this approximation we get IR equation
\begin{equation}
 \int d\tu_2\ G(\tu_1,\tu_2) \Sigma(\tu_2,\tu_3) = - \delta(\tu_1,\tu_3), \qquad \Sigma(\tu,\tu') = J^2\ G(\tu,\tu')^{3}
\end{equation}
it is easy to see that above equation are symmetric under following conformal transformations
\begin{equation}
 G\left(\tau, \tau^{\prime}\right) \rightarrow\left[f^{\prime}(\tau) f^{\prime}\left(\tau^{\prime}\right)\right]^{1/4} G\left(f(\tau), f\left(\tau^{\prime}\right)\right), \quad \Sigma\left(\tau, \tau^{\prime}\right) \rightarrow\left[f^{\prime}(\tau) f^{\prime}\left(\tau^{\prime}\right)\right]^{3/4} \Sigma\left(f(\tau), f\left(\tau^{\prime}\right)\right)
\end{equation}
the solutions to these equations at inverse temperature $\beta$ are 
\begin{equation}
 \Braket{\f{1}{N} \sum_i \psi_i(\tu_1)\psi_i(\tu_2)}_\beta = \braket{\psi_i(\tu_1)\psi_i(\tu_2)}_\beta = G^{\varphi}_\beta(\tu_1,\tu_2) = \f{C_\Delta}{\left[\f{J\beta}{\pi}\ \sin\left(\f{\pi}{\beta} \ (\varphi(\tu_1)-\varphi(\tu_2)\right)\right]^{2\Delta}}, 
\end{equation}
here $\Delta = \f{1}{4}$ and  $C_{\Delta} = \left[({1}/{4\pi})\tan\f{\pi}{4}\right]^\Delta$. 
The function $\varphi(\tu)$ is $S^1 \to S^1$ function where $\tu$ and $\varphi(\tu)$ both take values between $(-l',l)$. The solution to IR equations of motion parameterized by $\varphi \in (Diff\ S^1)/SL(2R)$ form the soft mode manifold. For the operator insertions in \eq{znp} we have to identify  $\psi_i(\tu_1)\psi_i(\tu_2)$ (no sum) with $G^{\varphi}_\beta(\tu_1,\tu_2)$. The string of operators inserted is
\begin{align}
 &\quad \sum_{k_j}^{N/2}\ \prod_{j=1}^{n}\ (-4)\ (s'_{k_j}s_{k_j})\ \psi_{2k_j-1}(\tu_j) \psi_{2k_j}(\tu_j)\ \psi^{2k_j-1}(-l') \psi^{2k_j}(-l')\ \psi_a(\tu_1)\psi_a(\tu_2)\hspace{6cm}\nn \\
 &=\sum_{k_j}^{N/2}\ (4)^n \left[(s'_{k_1}s_{k_1})\cdots (s'_{k_n}s_{k_n})\right]\ \left[{{\psi_{2k_1-1}}(\tu_1){\psi_{2k_1-1}}}(-l')\ {{\psi_{2k_1}}(\tu_1){\psi_{2k_1}}}(-l') \cdots \right. \nn \\ &\hspace{5.5cm} \cdots \left. {{\psi_{2k_n-1}}(\tu_n){\psi_{2k_n-1}}}(-l')\ {{\psi_{2k_n}}(\tu_n){\psi_{2k_n}(-l')}}\right]\ \psi_a(\tu_1)\psi_a(\tu_2)
\end{align}
the combination that is leading order in $N$ is
\begin{align}
&= \sum_{k_j}^{N/2}\ (4)^n \left[(s'_{k_1}s_{k_1})\cdots (s'_{k_n}s_{k_n})\right]\ \left[\wick[offset=1.5em]{\c1{\psi^{2k_1-1}}(\tu_1)\c1{\psi^{2k_1-1}(-l')}}\ \wick[offset=1.2em]{\c2{\psi^{2k_1}}(\tu_1)\c2{\psi^{2k_1}(-l')}} \cdots \hspace{3cm} \right. \nn \\ 
&\hspace{4cm}\cdots  \left. \wick[offset=1em]{\c3{\psi^{2k_n-1}}(\tu_n)\c3{\psi^{2k_n-1}(-l')}}\ \wick[offset=.8em]{\c4{\psi^{2k_n}}(\tu_n)\c4{\psi^{2k_n}(-l')}}\right]\ \wick[offset=.8em]{\c5{\psi_a}(\tu)\c5{\psi_a}}(\tu') + \cdots  \nn \\
& =G^{\varphi}_\beta(\tu_1,\tu_2) \prod_{j=1}^n\ 4 (s'.s)\ G^{\varphi}_\beta(\tu_j ,- l')^{2} \left(1 + O\left(\f{1}{N}\right)\right)\nn \\
& =N^n\ \left(G^{\varphi}_\beta(\tu_1,\tu_2)\prod_{j=1}^{n}\ \left[2\ \cos{\theta}\ G^{\varphi}_\beta(\tu_j ,- l')^{2}\right]\right) + O(N^{n-1}),
\end{align}
We have substituted $s'.s = (N/2) \cos\theta$. The saddle point approximation to \eq{znp} would be obtained by summing over all the saddles which would be an integral over the manifold of soft modes parameterized by $\varphi \in (Diff\ S^1)/SL(2R)$ and all of them have the same action in the conformal limit.
\begin{equation}
 \bar{\mathcal{A}}_n =\f{1}{\mathcal{Z}(\beta=l+l')} \oint \mathcal{D}\varphi\ e^{-S_0}\  
 N^n\ G^{\varphi}_\beta(\tu_1,\tu_2)\prod_{j=1}^{n}\ \left[2\ \cos{\theta}\ G^{\varphi}_\beta(\tu_j ,- l')^{2}\right].
\end{equation}
Here $S_0$ is the value conformal action on the soft mode manifold and $\mathcal{D}\varphi$ is an appropriate measure. We immediately see that this integral is divergent, to remedy this we do the lowest order perturbation in the operator $\del_\tau$ which would break the conformal symmetry and we will get an effective action for the soft modes, which is the famous Schwarzian action. Then in the slightly broken conformal limit we have
\begin{equation}
 \bar{\mathcal{A}}_n =\f{1}{\mathcal{Z}(\beta)} \oint \mathcal{D}\varphi\ \exp\left({\f{N \alpha_s}{J}\int d\tu\ \Big\{\tan\left(\f{\pi\varphi(\tu)}{\beta}\right),\ \tau \Big\}}\right)\  
 N^n\ G^{\varphi}_\beta(\tu_1,\tu_2)\prod_{j=1}^{n}\ \left[2\ \cos{\theta}\ G^{\varphi}_\beta(\tu_j ,- l')^{2}\right]
\end{equation}
Let us now plug this back into Euclidean version of \eq{fin} (this just removes the factor $i^n$ in \eq{fin} and replace $z \to \tau$, we will actually get the annealed averaged version of \eq{2pt})
\begin{align}
 \overline{G(t,t',l,l')} &=\f{1}{\mathcal{Z}(\beta)} \oint \mathcal{D}\varphi\ e^{{\f{N \alpha_s}{J}\int d\tu\ \big\{\tan\left(\f{\pi\varphi(\tu)}{\beta}\right),\ \tau \big\}}}\  
 \left[\sum_{n=0}^\infty \f{1}{n!}\ \left(N J \cos\theta\ \int d\tau \ep(\tau)\ G^{\varphi}_\beta(\tu ,- l')^{2}\right)^n\right]G^{\varphi}_\beta(\tu_1,\tu_2) \nn \\
 &=\f{1}{\mathcal{Z}(\beta)} \oint \mathcal{D}\varphi\quad e^{-S^E_{eff}[\varphi]}\quad G^{\varphi}_\beta(\tu_1,\tu_2)
\end{align}
where
\begin{equation}\label{seff}
 S^E_{eff}[\varphi] =- \f{N \alpha_s}{J}\left(\int d\tu\ \Big\{\tan\left(\f{\pi\varphi(\tu)}{\beta}\right),\ \tau \Big\}\ - \f{J^2\cos\theta}{\alpha_s}\ \int d\tau\ \ep(\tau)\ G^{\varphi}_\beta(\tu ,- l')^{2}\right).
\end{equation}
The superscript $E$ denotes that it is an Euclidean action.

\subsubsection*{Analytic continuation}
We have a path integral along the contour given by \autoref{c} and \eq{seff} is completely Euclidean  but as prescribed earlier while doing path integral along CTP we should analytically continue the action to the Lorentzian signature. Let us look at it more explicitly, we will divide the path integral in three parts from $z \in (il',0)$, $z \in (CTP)$ and $z \in (0,-il)$ (see \autoref{c}), then
\begin{equation}
 \overline{G(t,t',l,l')} = \f{1}{Z(\beta)} \int_{z \in (il',0)} \mathcal{D}\varphi\ e^{-S^{E_1}_{eff}[\varphi]} \quad \int_{ctp} \mathcal{D}\varphi \ e^{i S^L_{eff}[\varphi]} \quad \int_{z \in (0,-il)} \mathcal{D}\varphi\ e^{-S^{E_1}_{eff}[\varphi]}\ G^\varphi_\beta(it,it')
\end{equation}

where 
\begin{equation}
 S^{E_1}_{eff}[\varphi] =- \f{N \alpha_s}{J} \int d\tu\ \Big\{\tan\left(\f{\pi\varphi(\tu)}{\beta}\right),\ \tau \Big\}
\end{equation}
and
\begin{equation}
 S^L_{eff} = -\f{N \alpha_s}{J} \int_{ctp} dt\ \Big\{ \tanh\left(\f{\pi \varphi(t)}{\beta}\right),t \Big\} + N J \cos\theta\  C^2_{\Delta} \int_{ctp} dt\ \ep(t) \left[\f{J \beta}{\pi}\ \sin\left(\f{\pi(i\varphi(t) + l')}{\beta}\right) \right]^{-4\Delta}.
\end{equation}
When we take $l = l' = \beta/2$, then we get
\begin{equation}
  S^L_{eff} = -\f{N \alpha_s}{J} \int_{ctp} dt\ \Big\{ \tanh\left(\f{\pi \varphi(t)}{\beta}\right),t \Big\} + N J \cos\theta\  C^2_{\Delta} \int_{ctp} dt\ \ep(t) \left[\f{J \beta}{\pi}\ \cosh\left(\f{\pi \varphi(t)}{\beta}\right) \right]^{-4\Delta}.
\end{equation}
Now in the parameterization $f(t) = \f{\pi}{\beta J^2} \tanh\left(\f{\pi \varphi(t)}{\beta}\right)$
\begin{equation}
 S^L_{eff} = -\f{N \alpha_s}{J} \int_{ctp} dt\  \left[\{f(t),t\} - \f{J^2}{2} \left(\f{2\cos\theta C^2_{\Delta}}{\alpha_s}\right)\ \ep(t)\ (f')^{2\Delta}\right].
\end{equation}
Now, we define $\hat \ep(t) = ({2\cos\theta C^2_{\Delta}} /{\alpha_s})\ \ep(t)$ and that gives us  
\begin{equation}\label{eff}\boxed{
 S[f]:= S^L_{eff} = -\f{N \alpha_s}{J} \int_{ctp} dt\  \left[\{f(t),t\} - \f{J^2\hat \ep(t)}{2}\ (f')^{2\Delta}\right]}.
\end{equation}
Hence low energy effective action along the CTP is \eq{eff}.
\bibliographystyle{JHEP} \bibliography{1g}

\end{document}